\documentclass[aps,prd,preprint,superscriptaddress,showpacs,amsmath,amssym,letterpaper]{revtex4}

\usepackage{graphicx}
\usepackage{bm}
\usepackage{epsfig}

\RequirePackage{ifpdf}

\ifpdf
 \DeclareGraphicsExtensions{.jpg,.pdf,.jpeg}
\else
 \DeclareGraphicsExtensions{.eps,.ps,.eps.gz,.ps.gz}
\fi

\newcommand{\ppbar}{\ensuremath{p\bar{p}}}
\newcommand{\ttbar}{\ensuremath{t\bar{t}}}
\newcommand{\etal}{{\it et al.}}
\def \metnice {\ensuremath{{\not}{E_T}}}

\DeclareMathOperator{\JES}{JES}

\DeclareMathOperator{\bg}{bg}
\DeclareMathOperator{\eff}{eff}
\DeclareMathOperator{\tree}{tree}
\newcommand{\LtwoD}{L_{\textrm{2D}}}

\begin{document}

\title{Top Quark Mass Measurement in the Lepton plus Jets Channel Using 
    a~Modified Matrix Element Method }

\affiliation{Institute of Physics, Academia Sinica, Taipei, Taiwan 11529, Republic of China} 
\affiliation{Argonne National Laboratory, Argonne, Illinois 60439} 
\affiliation{University of Athens, 157 71 Athens, Greece} 
\affiliation{Institut de Fisica d'Altes Energies, Universitat Autonoma de Barcelona, E-08193, Bellaterra (Barcelona), Spain} 
\affiliation{Baylor University, Waco, Texas  76798} 
\affiliation{Istituto Nazionale di Fisica Nucleare Bologna, $^v$University of Bologna, I-40127 Bologna, Italy} 
\affiliation{Brandeis University, Waltham, Massachusetts 02254} 
\affiliation{University of California, Davis, Davis, California  95616} 
\affiliation{University of California, Los Angeles, Los Angeles, California  90024} 
\affiliation{University of California, San Diego, La Jolla, California  92093} 
\affiliation{University of California, Santa Barbara, Santa Barbara, California 93106} 
\affiliation{Instituto de Fisica de Cantabria, CSIC-University of Cantabria, 39005 Santander, Spain} 
\affiliation{Carnegie Mellon University, Pittsburgh, PA  15213} 
\affiliation{Enrico Fermi Institute, University of Chicago, Chicago, Illinois 60637}
\affiliation{Comenius University, 842 48 Bratislava, Slovakia; Institute of Experimental Physics, 040 01 Kosice, Slovakia} 
\affiliation{Joint Institute for Nuclear Research, RU-141980 Dubna, Russia} 
\affiliation{Duke University, Durham, North Carolina  27708} 
\affiliation{Fermi National Accelerator Laboratory, Batavia, Illinois 60510} 
\affiliation{University of Florida, Gainesville, Florida  32611} 
\affiliation{Laboratori Nazionali di Frascati, Istituto Nazionale di Fisica Nucleare, I-00044 Frascati, Italy} 
\affiliation{University of Geneva, CH-1211 Geneva 4, Switzerland} 
\affiliation{Glasgow University, Glasgow G12 8QQ, United Kingdom} 
\affiliation{Harvard University, Cambridge, Massachusetts 02138} 
\affiliation{Division of High Energy Physics, Department of Physics, University of Helsinki and Helsinki Institute of Physics, FIN-00014, Helsinki, Finland} 
\affiliation{University of Illinois, Urbana, Illinois 61801} 
\affiliation{The Johns Hopkins University, Baltimore, Maryland 21218} 
\affiliation{Institut f\"{u}r Experimentelle Kernphysik, Universit\"{a}t Karlsruhe, 76128 Karlsruhe, Germany} 
\affiliation{Center for High Energy Physics: Kyungpook National University, Daegu 702-701, Korea; Seoul National University, Seoul 151-742, Korea; Sungkyunkwan University, Suwon 440-746, Korea; Korea Institute of Science and Technology Information, Daejeon, 305-806, Korea; Chonnam National University, Gwangju, 500-757, Korea} 
\affiliation{Ernest Orlando Lawrence Berkeley National Laboratory, Berkeley, California 94720} 
\affiliation{University of Liverpool, Liverpool L69 7ZE, United Kingdom} 
\affiliation{University College London, London WC1E 6BT, United Kingdom} 
\affiliation{Centro de Investigaciones Energeticas Medioambientales y Tecnologicas, E-28040 Madrid, Spain} 
\affiliation{Massachusetts Institute of Technology, Cambridge, Massachusetts  02139} 
\affiliation{Institute of Particle Physics: McGill University, Montr\'{e}al, Qu\'{e}bec, Canada H3A~2T8; Simon Fraser University, Burnaby, British Columbia, Canada V5A~1S6; University of Toronto, Toronto, Ontario, Canada M5S~1A7; and TRIUMF, Vancouver, British Columbia, Canada V6T~2A3} 
\affiliation{University of Michigan, Ann Arbor, Michigan 48109} 
\affiliation{Michigan State University, East Lansing, Michigan  48824}
\affiliation{Institution for Theoretical and Experimental Physics, ITEP, Moscow 117259, Russia} 
\affiliation{University of New Mexico, Albuquerque, New Mexico 87131} 
\affiliation{Northwestern University, Evanston, Illinois  60208} 
\affiliation{The Ohio State University, Columbus, Ohio  43210} 
\affiliation{Okayama University, Okayama 700-8530, Japan} 
\affiliation{Osaka City University, Osaka 588, Japan} 
\affiliation{University of Oxford, Oxford OX1 3RH, United Kingdom} 
\affiliation{Istituto Nazionale di Fisica Nucleare, Sezione di Padova-Trento, $^w$University of Padova, I-35131 Padova, Italy} 
\affiliation{LPNHE, Universite Pierre et Marie Curie/IN2P3-CNRS, UMR7585, Paris, F-75252 France} 
\affiliation{University of Pennsylvania, Philadelphia, Pennsylvania 19104}
\affiliation{Istituto Nazionale di Fisica Nucleare Pisa, $^x$University of Pisa, $^y$University of Siena and $^z$Scuola Normale Superiore, I-56127 Pisa, Italy} 
\affiliation{University of Pittsburgh, Pittsburgh, Pennsylvania 15260} 
\affiliation{Purdue University, West Lafayette, Indiana 47907} 
\affiliation{University of Rochester, Rochester, New York 14627} 
\affiliation{The Rockefeller University, New York, New York 10021} 
\affiliation{Istituto Nazionale di Fisica Nucleare, Sezione di Roma 1, $^{aa}$Sapienza Universit\`{a} di Roma, I-00185 Roma, Italy} 

\affiliation{Rutgers University, Piscataway, New Jersey 08855} 
\affiliation{Texas A\&M University, College Station, Texas 77843} 
\affiliation{Istituto Nazionale di Fisica Nucleare Trieste/Udine, I-34100 Trieste, $^{bb}$University of Trieste/Udine, I-33100 Udine, Italy} 
\affiliation{University of Tsukuba, Tsukuba, Ibaraki 305, Japan} 
\affiliation{Tufts University, Medford, Massachusetts 02155} 
\affiliation{Waseda University, Tokyo 169, Japan} 
\affiliation{Wayne State University, Detroit, Michigan  48201} 
\affiliation{University of Wisconsin, Madison, Wisconsin 53706} 
\affiliation{Yale University, New Haven, Connecticut 06520} 
\author{T.~Aaltonen}
\affiliation{Division of High Energy Physics, Department of Physics, University of Helsinki and Helsinki Institute of Physics, FIN-00014, Helsinki, Finland}
\author{J.~Adelman}
\affiliation{Enrico Fermi Institute, University of Chicago, Chicago, Illinois 60637}
\author{T.~Akimoto}
\affiliation{University of Tsukuba, Tsukuba, Ibaraki 305, Japan}
\author{B.~\'{A}lvarez~Gonz\'{a}lez$^q$}
\affiliation{Instituto de Fisica de Cantabria, CSIC-University of Cantabria, 39005 Santander, Spain}
\author{S.~Amerio$^w$}
\affiliation{Istituto Nazionale di Fisica Nucleare, Sezione di Padova-Trento, $^w$University of Padova, I-35131 Padova, Italy} 

\author{D.~Amidei}
\affiliation{University of Michigan, Ann Arbor, Michigan 48109}
\author{A.~Anastassov}
\affiliation{Northwestern University, Evanston, Illinois  60208}
\author{A.~Annovi}
\affiliation{Laboratori Nazionali di Frascati, Istituto Nazionale di Fisica Nucleare, I-00044 Frascati, Italy}
\author{J.~Antos}
\affiliation{Comenius University, 842 48 Bratislava, Slovakia; Institute of Experimental Physics, 040 01 Kosice, Slovakia}
\author{G.~Apollinari}
\affiliation{Fermi National Accelerator Laboratory, Batavia, Illinois 60510}
% ...

\collaboration{CDF Collaboration\footnote{With visitors from $^a$University of Massachusetts Amherst, Amherst, Massachusetts 01003,
$^b$Universiteit Antwerpen, B-2610 Antwerp, Belgium, 
$^c$University of Bristol, Bristol BS8 1TL, United Kingdom,
$^d$Chinese Academy of Sciences, Beijing 100864, China, 
$^e$Istituto Nazionale di Fisica Nucleare, Sezione di Cagliari, 09042 Monserrato (Cagliari), Italy,
$^f$University of California Irvine, Irvine, CA  92697, 
$^g$University of California Santa Cruz, Santa Cruz, CA  95064, 
$^h$Cornell University, Ithaca, NY  14853, 
$^i$University of Cyprus, Nicosia CY-1678, Cyprus, 
$^j$University College Dublin, Dublin 4, Ireland,
$^k$University of Edinburgh, Edinburgh EH9 3JZ, United Kingdom, 
$^l$Universidad Iberoamericana, Mexico D.F., Mexico,
$^m$Queen Mary, University of London, London, E1 4NS, England,
$^n$University of Manchester, Manchester M13 9PL, England, 
$^o$Nagasaki Institute of Applied Science, Nagasaki, Japan, 
$^p$University of Notre Dame, Notre Dame, IN 46556,
$^q$University de Oviedo, E-33007 Oviedo, Spain, 
$^r$Texas Tech University, Lubbock, TX  79409, 
$^s$IFIC(CSIC-Universitat de Valencia), 46071 Valencia, Spain,
$^t$University of Virginia, Charlottesville, VA  22904,
$^u$Bergische Universit\"at Wuppertal, 42097 Wuppertal, Germany,
$^{cc}$On leave from J.~Stefan Institute, Ljubljana, Slovenia, 
}}
\noaffiliation

\date{\today}

\begin{abstract}
We report a measurement of the top quark mass, $m_t$, obtained from
$\ppbar$ collisions at $\sqrt{s} = 1.96$~TeV at the Fermilab Tevatron
using the CDF II detector.  We analyze a sample corresponding to an
integrated luminosity of~1.9~fb$^{-1}$. We select events with an
electron or muon, large missing transverse energy, and exactly four
high-energy jets in the central region of the detector, at least one
of which is tagged as coming from a $b$ quark.  We calculate a signal
likelihood using a matrix element integration method, where the matrix
element is modified by using effective propagators to take into
account assumptions on event kinematics.  Our event likelihood is
a~function of $m_t$ and a parameter JES that determines {\it{in situ}}
the calibration of the jet energies.  We use a neural network
discriminant to distinguish signal from background events. We also
apply a cut on the peak value of each event likelihood curve to reduce
the contribution of background and badly reconstructed events.  Using
the 318 events that pass all selection criteria, we find $m_t = 172.7
\pm 1.8$ (stat. + JES) $\pm$ 1.2 (syst.) GeV/$c^2$.

\end{abstract}


\maketitle

\section{Introduction}
\label{sec:intro}

The top quark mass, $m_t$, is an important parameter in the standard
model of particle physics. Since the discovery of the top quark in
1995, there have been many reported measurements of its mass, all from
the CDF and D0~experiments at the Fermilab Tevatron~\cite{TevEW}. The
standard model relates the top quark and $W$ boson masses to the mass
of the predicted Higgs boson via loop corrections. Precision
measurements of $m_t$ and the $W$ boson mass $m_W$, in conjunction
with many other precision electroweak measurements, thus provide
constraints on the value of the Higgs boson mass~\cite{LEPcollab}.

The measurement reported here uses $\ppbar$ collision data
corresponding to an integrated luminosity of 1.9 fb$^{-1}$, collected
by the CDF II detector during Run II of the Fermilab Tevatron collider
at $\sqrt{s} = 1.96$~TeV. In $\ppbar$ collisions, top quarks are
produced predominantly as $\ttbar$ pairs, and present measurements
within the standard model framework indicate that the top quark decays
to a $W$ boson and a $b$ quark nearly 100\% of the
time~\cite{pdg}. The $W$ boson can decay into either a charged lepton
and a~neutrino (``leptonic decay'') or a quark-antiquark pair
(``hadronic decay''). We select events in which one of the $W$ bosons
decays leptonically and the other decays hadronically, where the
lepton in the leptonic decay is required to be an electron or muon;
this channel is referred to as the ``lepton + jets'' channel.  Decays
of $W$ bosons into a tau lepton are not explicitly included in our
model, although some events containing a tau lepton which decays into
an electron or muon do pass our selection criteria and amount to
approximately 7\% of the $\ttbar$ signal.  In general, an event in our
candidate sample has four high-energy jets (two of which come from the
parton shower and hadronization associated with the quarks from the
hadronic $W$ boson decay and two from the parton shower and
hadronization of the $b$ quarks), a charged electron or muon, and an
unobserved neutrino. For a given Tevatron integrated luminosity, the
lepton + jets channel allows for more precise measurements than
channels in which both $W$ bosons decay leptonically or hadronically,
as it offers the best balance of available statistics and sample
purity.  The most recent $m_t$ measurements obtained at the Tevatron
using the lepton + jets topology are reported in Ref.~\cite{ljettev}.

The method we use to extract the top quark mass from a sample of
candidate $\ttbar$ events is a modified matrix element integration
method. The matrix element approach to the top quark mass
measurement~\cite{Kondo-Estrada} is based on integrating
over the tree-level phase space of the process, where each kinematic
configuration of the tree-level partons is weighted by the matrix
element squared and by the probability that the detector observables
can be produced by the final state particles.
With an appropriate normalization factor, this integral defines
the probability to see an event with this configuration in the detector.
By multiplying the individual
event probabilities, we obtain a likelihood, as a function of $m_t$, of
seeing the event sample observed in our detector.

In theory, the distributions of the invariant masses of the top quark
and the $W$ boson decay products are dominated by propagator-induced
terms in the matrix element squared. The top quark and the $W$ boson
widths are relatively narrow in comparison with their respective
masses. This leads to nearly pure relativistic Breit-Wigner
distributions for the invariant masses of their decay
products. Compared to this theoretical prediction, finite resolution
of the detector measurement naturally results in a widening of the
observed distributions. We describe the widening due to an imperfect
measurement of magnitudes of jet momenta in terms of detector transfer
functions. Effects of other uncertainties, such as finite angular
resolutions, are modeled by replacing the Breit-Wigner terms in the matrix
element squared with empirically determined distributions called
``effective propagators'' in this paper. This modification of the
matrix element improves the observation model for $\ttbar$ events.

The matrix element used in this work~\cite{kleiss} includes $\ttbar$
production, from both quark-antiquark and gluon-gluon collisions, and
decay into the lepton + jets channel. Since we do not know which jet
observed in our detector corresponds to which parton in the matrix
element, we calculate the likelihood for each possible assignment of
jets to partons and sum the likelihood over all permutations. Each
permutation includes a weight, which takes into account the
probability that the permutation is consistent with the observed
information on whether the jet has been tagged or not as a
$b$-jet. Tagging of $b$-jets is done by the displaced vertex technique
discussed in Section~\ref{sec:sample}.  For each permutation, the
matrix element integration is performed over the seven kinematic
variables in the event that remain after a set of simplifying
assumptions.

We improve the precision of the method by introducing another
parameter into our likelihood, the jet energy scale (JES). This is a
scale factor which multiplies the energy of all jets. The uncertainty
on the jet energy scale is the
major source of systematic uncertainty on the top mass measurement;
by including JES as a parameter in the likelihood, we can use the
$W$ boson decay to hadrons in the $\ttbar$ decay chain to 
provide {\it in situ} calibration, thus reducing the systematic
uncertainty due to JES. Our final likelihood calculated
for each candidate event is thus a function of both the
top quark mass and JES.

Our model is designed to fit lepton+jets $\ttbar$ events where the
final objects observed in the event come directly from $\ttbar$
decays. We thus need to take into account non-$\ttbar$ events or
$\ttbar$ events where some of the observed objects do not originate
from $\ttbar$ decay. The probability that a $\ttbar$ candidate event
is background is estimated using a neural network output that is a
function of several shape and kinematic variables, and then their
expected contribution to the likelihood is subtracted from the total
likelihood. In addition, we cut on the magnitude of the peak of the
likelihood for an event to further reduce background and badly modeled
events.

We multiply the individual likelihood curves from each event to get an
overall likelihood. Because of the assumptions made, and the presence
of background events, the extraction of a mass value and its
uncertainty needs a calibration, which we obtain from Monte Carlo
simulated events.

Section~\ref{sec:detect} is a brief description of the CDF II detector and
its use for the measurements needed in this
analysis. Section~\ref{sec:sample} defines the data sample used for
this analysis and the estimated background.
Section~\ref{sec:analysis} describes the likelihood construction
 for $\ttbar$ signal events.  Section~\ref{sec:tot-likeli}
explains how non-$\ttbar$ events are
incorporated into the likelihood function. Section~\ref{sec:calib}
describes how the method is tested and calibrated. Section~\ref{sec:systs}
covers the systematic uncertainties.
Section~\ref{sec:result} summarizes the results obtained by applying the
method to the data. Finally, Section~\ref{sec:summary} gives the
conclusions.

\section{The CDF II Detector}
\label{sec:detect}
A complete description of the CDF II detector and its use in lepton,
jet, and secondary vertex reconstruction can be found
elsewhere~\cite{CDFII}. Here, we describe the components that are
essential for this analysis and how they are used.

The CDF II detector is a general-purpose detector with a cylindrical
geometry featuring forward-backward symmetry and axial symmetry around
the beam pipe. The CDF coordinate system uses a cylindrical system
centered at the interaction point with the $z$ (longitudinal) axis
along the proton beam direction, $r$ the distance to the beam line,
and $\phi$ the azimuthal angle around the beam line. We also use
$\theta$, the polar angle from the beam line. The pseudorapidity
$\eta$ of a particle three-momentum is defined in terms of the polar
angle $\theta$ by $\eta = -\ln(\tan(\theta/2))$. For a particle with
momentum $p$ and energy $E$, we define the transverse momentum $p_T$
and the transverse energy $E_T$ as $p \sin \theta$ and $E \sin
\theta$, respectively. The detector covers the complete solid angle in
$\theta$ and $\phi$ up to $|\eta| = 3.6$.

The innermost part of the detector consists of the charged particle
tracking detectors, which are immersed in a 1.4 T magnetic field
provided by a superconducting solenoid oriented parallel to the beam
axis. Calorimeters and muon systems outside the solenoid provide
lepton measurement and identification in addition to jet momentum
measurements. The tracking detectors and calorimeters together provide
identification of jets from heavy (charm and bottom) quarks.

 
The first component of the tracking system is a series of silicon
microstrip detectors between radii of 1.5 and 28 cm. The innermost
layer (L00)~\cite{l00} is a single-sided layer of silicon attached
directly to the beam pipe, providing a position measurement very close
to the collision point. Five layers of double-sided microstrip
detectors (SVXII) cover up to $r = 10.6$ cm in the $|\eta|<$ 1.0
region~\cite{svx2}. Each layer has one side with strips oriented
parallel to the beam axis to provide measurements in the $r$--$\phi$
plane and one side at a stereo angle to provide three-dimensional
measurements; two layers have strips at a $90^{\circ}$ angle and three
layers have strips at a $1.2^{\circ}$ angle. The ISL~\cite{isl} is an
additional set of silicon microstrip detectors located outside of SVXII
to provide measurements at larger distances from the beam line, thus 
improving the silicon tracking. It consists
of one layer at $r = 22$ cm in the central region ($|\eta|
< 1.0$) and two layers at $r = 20$ cm and $r = 28$ cm in the forward
region ($1.0 < |\eta| < 2.0$).
The typical resolution of these detectors in the $r$--$\phi$ plane is
11 $\mu$m. The impact parameter resolution of this system is
$\sigma(d_0)
\approx 40~\mu$m, of which approximately $35~\mu$m is due to the transverse
size of the Tevatron interaction region. Outside of the silicon layers
lies the central outer tracker (COT)~\cite{cot}, an open-cell drift
chamber detector, which provides coverage for $|\eta|<$ 1.0. Multiple
wire planes, each with 12 sense wires, are grouped in 8 superlayers
which extend to a radius of 137 cm. The superlayers alternate between
having wires parallel to the beam axis and wires skewed by a $\pm
2^\circ$ stereo angle, thus providing up to 96 points for track
reconstruction. Together with the additional constraint coming from
the primary vertex position, these tracking elements provide 
a resolution on the track transverse momentum, 
$p_T$, of $\sigma(p_T)/p_T \approx 0.1\% \cdot p_T/(\mathrm{GeV}/c)$.

Outside the tracking system and the solenoid are segmented
electromagnetic and hadronic calorimeters. The central calorimeter
covers up to $|\eta|<$ 1.1 and has a projective geometry consisting of
towers segmented in $\eta$ and $\phi$ pointing toward the center of
the detector. The central electromagnetic calorimeter (CEM) consists
of alternating layers of lead plates and plastic scintillators, 18
radiation lengths deep. The energy detected in small contiguous groups
of calorimeter towers is summed into electromagnetic clusters. These
clusters are identified as electron candidates if they match a track
reconstructed in the tracking system and if very little energy is
detected in the surrounding towers (i.e., if the cluster is
isolated). The electron energy resolution for an electron with
tranverse energy $E_T$ is given by $\sigma(E_T)/E_T \approx
13.5\%/\sqrt{E_T/\mathrm{GeV}}~\oplus$ 2\%.  Approximately at shower
maximum are proportional strip and wire chambers (CES) which provide
finer position resolution for electron and photon identification. The
central hadronic calorimeter (CHA) is composed of alternating layers
of iron plates and scintillators, 4.5 nuclear interaction lengths
deep, again with projective geometry segmentation. A plug tile
calorimeter covers the forward region with $1.1 < |\eta| < 3.6$,
consisting of a lead/scintillator electromagnetic portion (PEM),
scintillator strips at shower maximum (PES), and an iron/scintillator
hadronic portion (PHA). An additional hadronic calorimeter (WHA)
covers the region between the plug calorimeter and the central
calorimeter and improves the hermeticity of the detector.  These
calorimeters provide jet measurements with a resolution of
approximately $\sigma(E_T) \approx 0.1 \cdot E_T +1.0$ GeV~\cite{jetres}.

In the central and forward regions, jets are reconstructed with a cone
algorithm~\cite{jetres}, which adds groups of electromagnetic
($E_\mathrm{EM}$) and hadronic clusters ($E_\mathrm{HAD}$) that fall
within a cone of radius $\Delta R = \sqrt{\Delta \phi^2 + \Delta
\eta^2} \le 0.4$ around a seed tower with energy of at least 1
GeV. The jet energies are corrected for  multiple
primary interactions (pileup) and for detector effects including a 
calibrated non-linearity in the calorimeter
and average losses in non-sensitive regions of the calorimeter. 
The jet energies are also corrected for hadronic physics effects.  
Soft hadroproduction in the 
underlying event tends to increase the measured jet energy, while the 
limited cone size of the jet clustering algorithm gives rise to
out-of-cone losses~\cite{nimjets}. 
Uncertainties for each of these corrections contribute to the
jet systematic uncertainty, and are used to assign an uncertainty
on the top quark mass.

In the lepton + jets channel one of the $W$ bosons decays into a
lepton and a neutrino, which escapes undetected. This results in less
energy being measured in our detector than we would otherwise
expect. We require this as a signature for $\ttbar$
events. Specifically, we define a quantity, the missing $E_T$
($\metnice$), to measure the resulting transverse energy imbalance as follows:

\begin{eqnarray}
\metnice = |\sum_{i~\in~\mathrm{towers}}{E_{T_i} {\hat{n}_{T_i}}} |,
\end{eqnarray}

where $\hat{n}_{T_i}$ is the unit vector in the $x$--$y$ plane pointing 
from the primary vertex to  a given calorimeter tower and $E_{T_i}$ 
is the uncorrected 
$E_T$ measured in that tower. Two additional corrections to this
quantity are made. One is to account for muons, which, unlike other
particles, typically deposit only a small fraction of their energy in
the towers, and the other is to take into account the corrections
applied to the raw energies of the jets. Details of these corrections
can be found in Ref.~\cite{dlm}.


Muon identification takes place in three separate subdetectors. Two of
these are in the central region: one set of four layers of drift
chambers (CMU) located outside the central calorimeters (after 4.6
hadronic absorption lengths of material), and another set of four
layers (CMP) located outside the magnet return yoke, which provides an
additional 60 cm of absorbing steel. Muon tracks in this region are
required to pass through both detectors and are called CMUP
muons. These two subdetectors cover the region $|\eta| \le 0.6$. Muons in the
region 0.6 $<|\eta|<$ 1.0 are detected by an additional set of four
layers of drift chambers (CMX), completing the full fiducial region of
the COT. CMUP or CMX track segments are matched to tracks in the COT;
in addition, the energy deposited in the CEM and CHA is required to be
small.

The trigger system is used to record events with high-$p_T$
leptons. The trigger is a three-level filter in which the first two
levels use specialized hardware and utilize only the detector
subsystems with fast readout.  The third level is a complete
reconstruction of the event using the same software used for the
offline reconstruction, but with less stringent cuts.  The level~1
(L1) trigger uses information from the calorimeter clusters and from
the XFT (eXtremely Fast Tracker), which reconstructs tracks from the
COT $r$--$\phi$ information with a momentum resolution given by
$\sigma(p_T)/p_T \approx 2\% \cdot
p_T/(\mathrm{GeV}/c)$~\cite{xft}. The L1 central electron trigger requires a
track with $p_T > 8$ GeV/$c$ pointing to a tower with $E_T > 8$ GeV
and $E_\mathrm{HAD}/E_\mathrm{EM} < 0.125$. The L2 trigger adds
clustering in the CEM calorimeter and requires that a cluster with $E_T >
16$ GeV matches with a  $p_T > 8$ GeV/$c$ track. The L1
and L2 muon triggers require a track with $p_T > 4$ GeV/$c$ (CMUP) or
$p_T > 8$ GeV/$c$ (CMX) pointing to a track segment in the respective
drift chamber system. A complete lepton reconstruction is performed in
the L3 trigger, where $E_T > 18$ GeV is required for electrons and
$p_T > 18$ GeV/$c$ is required for muons.

\section{Data Sample and Background}
\label{sec:sample}

As mentioned previously, we search for events in the lepton + jets
topology, where a $\ttbar$ pair is produced, each top quark decays into a
$W$ boson and a $b$ quark, and one $W$ boson decays leptonically and
one hadronically.
We thus identify our top quark candidates by selecting events with four
high-energy jets, a~high-energy electron or muon, and $\metnice$ from
a~neutrino. Specifically, we require either an electron with $E_T > 20$
GeV or a muon with $p_T > 20$ GeV/$c$ in the central region ($|\eta|
< 1.0$) of the
detector. Electron and muon identification criteria are discussed in
Ref.~\cite{CDFII}. For the neutrino, we require $\metnice > 20$
GeV in the event. We require exactly 4 jets with $E_T > 20$ GeV
and pseudorapidity $|\eta| < 2.0$. The jet $E_T$ is corrected for pileup,
inhomogeneities of the detector,  and nonlinear calorimeter response as
a function of jet $p_T$ and $\eta$~\cite{nimjets}. The additional corrections
(underlying event and out-of-cone losses) are not used in the
analysis, but their uncertainties are taken into account in evaluating
the systematic uncertainties on the final result. 

Non-$\ttbar$ events that contain a $W$ boson and four jets are able to
pass the aforementioned selection cuts. However, most of these events
do not contain $b$ quarks in their final state, while $\ttbar$ events
will nearly always have two $b$ quarks. The $b$ quarks from top quark
decay hadronize into B-hadrons with energies on the order of several
tens of GeV, due to the high mass of the parent top quark. Since the B
hadron decay time is approximately 1.5 ps, it is possible to
reconstruct secondary vertices within a jet using the charged
particles from the B decay~\cite{secv-sigma}. A jet with an identified
secondary vertex is called a $b$-tagged jet. Therefore, to further
increase the $\ttbar$ purity of the sample, we require that at least
one of the jets must be tagged as a $b$-jet using a secondary vertex
tagging algorithm.

\begin{figure}[htbp]
\centerline {
            \epsfxsize 0.6\textwidth \epsffile{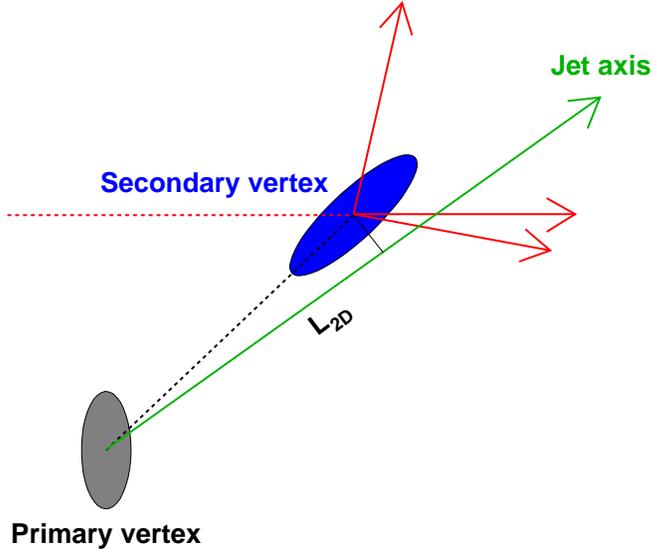} }
\caption{ \label{fig:svxtag} A view of a $b$-jet in the $x$--$y$
            plane. $\LtwoD$, the distance between the primary and
            secondary vertices projected onto the jet axis, along with
            its uncertainty, is used to determine whether a jet
            originates from a heavy flavor quark.}
\end{figure}

The outline of the $b$-tagging algorithm used, {\sc secvtx}, is as
follows: first, the charged particle tracks in the jet are subjected
to selection cuts to ensure that a quality secondary vertex can be
reconstructed. There must either be at least three tracks with $p_T
\ge 0.5$ GeV/$c$ where at least one of the tracks must be $\ge 1$
GeV/$c$, or at least two tracks with $p_T \ge 1$ GeV/$c$. Once the
secondary vertex is reconstructed using the tracks, the distance in
the $x$--$y$ plane between the primary and secondary vertices is
projected onto the direction of the jet; this quantity is referred to
as $\LtwoD$ (see Fig.~\ref{fig:svxtag}).  A jet is tagged if $\LtwoD
> 7.5~ \sigma_{\LtwoD}$, where $\sigma_{\LtwoD}$, the uncertainty on
$\LtwoD$, is approximately 190 $\mu$m. For $b$-jets produced in
$\ttbar$ decay, the $b$-tagging efficiency is about 40\%, while light
jets are misidentified as $b$-jets with a rate of less than 2\%.  For
more details see Ref.~\cite{secv-sigma}.

In 1.9 fb$^{-1}$ of data we find 371 $\ttbar$ candidate events that
pass the above selection requirements, 284 of which have one $b$-tag
and 87 of which have more than one $b$-tag (see Table~\ref{ev_back});
207 of the candidate events contain an electron and 164 contain a
muon.  The background to the $\ttbar$ signal consists of three main
sources: a) events where a $W$ boson is produced in conjunction with
heavy flavor quarks ($b\bar{b}$, $c\bar{c}$, or $c$); b) events where
a $W$ boson is produced along with light flavor quarks where a light
flavor jet has been incorrectly tagged with a $b$-tag (mistag); c) QCD
events, which do not contain a $W$ boson (non-$W$ events) but have a
jet mimicking a lepton, a jet with a $b$-tag, and $\metnice$. There
are also smaller contributions from single top quark production,
diboson ($WW$, $WZ$, or $ZZ$) production, and $Z$ + jets
production. The estimated number of background events for each of
these sources is derived with the method used for the cross section
measurement~\cite{mtwoforu}.

The contributions for the various types of background shown in
Table~\ref{ev_back} are estimated as follows. First, we define a
pretag event sample, which comprises all events that pass all the
signal selection requirements except for the $b$-tag requirement; our
final tagged samples are thus subsets of the pretag sample. For all
samples, the expected number of events for diboson, $Z$ + jets, and
single top quark backgrounds, as well as the $\ttbar$ signal, are
estimated using Monte Carlo (MC) simulated events assuming the
theoretical cross sections. To simulate the signal, we generate
$\ttbar$ events at a variety of top quark masses from 152 GeV/$c^2$ to
190 GeV/$c^2$ (needed for our top mass analysis) using the {\sc
pythia} MC generator version 6.216~\cite{pythia}. As a cross-check we
also use $\ttbar$ signal events generated with the {\sc herwig}
generator version 6.510~\cite{herwig}. For the number of expected
$\ttbar$ events used in the background estimate we use a top mass of
175 GeV/c$^2$, with a calculated $t\bar t$ production cross section of
$6.7 \pm 0.8$ pb~\cite{mlm}.

The non-$W$ contribution is estimated by a fit to the observed
$\metnice$ distribution of expected $\metnice$ distributions for
non-$W$ events (which lie mostly in the low $\metnice$ region) and
$W$+jets events. These distributions are taken from data sidebands
(either events with leptons which fail to meet the isolation
requirements, or from ``antielectron'' samples, which are electron
candidates failing two other selection requirements) and from
simulated MC events. This fit is performed separately for the
pretagged and tagged samples to obtain the expected number of events
in each.
  
The $W$ + jets background contribution
to the pretag sample is taken as the remainder after subtracting all the
above pretag contributions. The relative contribution of $W$ +
heavy flavor events to the pretag $W$ + jets contribution is estimated
with MC simulation. We use MC
events generated with the {\sc alpgen}~\cite{alpgen} generator
(version 2.10 prime) along with {\sc pythia} version 6.325 to perform
the parton shower and hadronization.
The {\sc alpgen} program is used
to generate samples with specific numbers of partons in the matrix element; 
this decreases
the time to generate events with high jet multiplicity. Since each
sample contains a different number of partons (for instance, the $W +
b\bar{b}$ sample contains $W + b\bar{b}+0p$, $W + b\bar{b}+1p$, and $W
+ b\bar{b}+\ge 2p$ contributions), we combine the separate samples
using their expected fractions, and remove overlaps using the {\sc
alpgen} jet-parton matching along with a jet-based heavy flavor
overlap removal algorithm~\cite{mtwoforu}.
Finally, applying heavy flavor and light flavor
$b$-tag efficiencies we obtain the estimated $W$ + jets contributions
to the final sample.

The single top quark contribution is generated using the
MadGraph/MadEvent~\cite{madgraph} package along with {\sc pythia} for
the parton shower and hadronization. Since their expected contribution
is small, we do not use separate MC samples for diboson or $Z$+jets
backgrounds, but rather merge them into the $W$ + light flavor
total. All MC samples are simulated using the CDF II detector response
simulation package~\cite{simulation}.

Table~\ref{ev_back} summarizes the data sample composition as a function of the
number of tagged jets in the event.
The total number of expected background events in our
data sample is $N_{\rm bg} = 70.3 \pm 16.5$  out of 371 observed
events. 

\begin{table} [htbp]
\caption{Summary of observed data and predicted $\ttbar$ signal and background contributions as a function of $b$-tags in the event.}
\label{ev_back}
\begin{center}
\begin{tabular}{l@{\hspace{1.0cm}}c@{\hspace{1.0cm}}c}
\hline \hline
Background                    & 1 tag  & $\ge$ 2 tags    \\
\hline
non-$W$ QCD                     & 13.8 $\pm$ 11.5  & 0.5 $\pm$ 1.5 \\
$W$ + light flavor (mistag)    & 16.3 $\pm$ 3.6   & 0.3 $\pm$ 0.1 \\
Diboson ($WW$, $WZ$, $ZZ$), $Z$ + jets & 5.5 $\pm$ 0.4 & 0.5 $\pm$ 0.1 \\
$W$ + $b\bar{b}$, $c\bar{c}$, $c$  & 26.1 $\pm$ 10.2 & 3.4 $\pm$ 1.4 \\
Single top                    & 3.0 $\pm$ 0.2      & 0.9 $\pm$ 0.1 \\
\hline 
Total background              & 64.7 $\pm$ 16.3  & 5.5 $\pm$ 2.6 \\
Predicted $\ttbar$ signal     & 182.6 $\pm$ 24.6 & 69.4 $\pm$ 11.2 \\
Events observed               & 284                &  87  \\
\hline \hline
\end{tabular}
\end{center}
\end{table}

We test our background model by comparing selected kinematic
distributions in the data with those expected by our model, by adding
the MC samples used for the $\ttbar$ signal and backgrounds and the
data samples used for the QCD background according to their predicted
proportions.

\begin{figure}[htbp]
\begin{center}
  \includegraphics*[width=0.9\textwidth,clip=true]{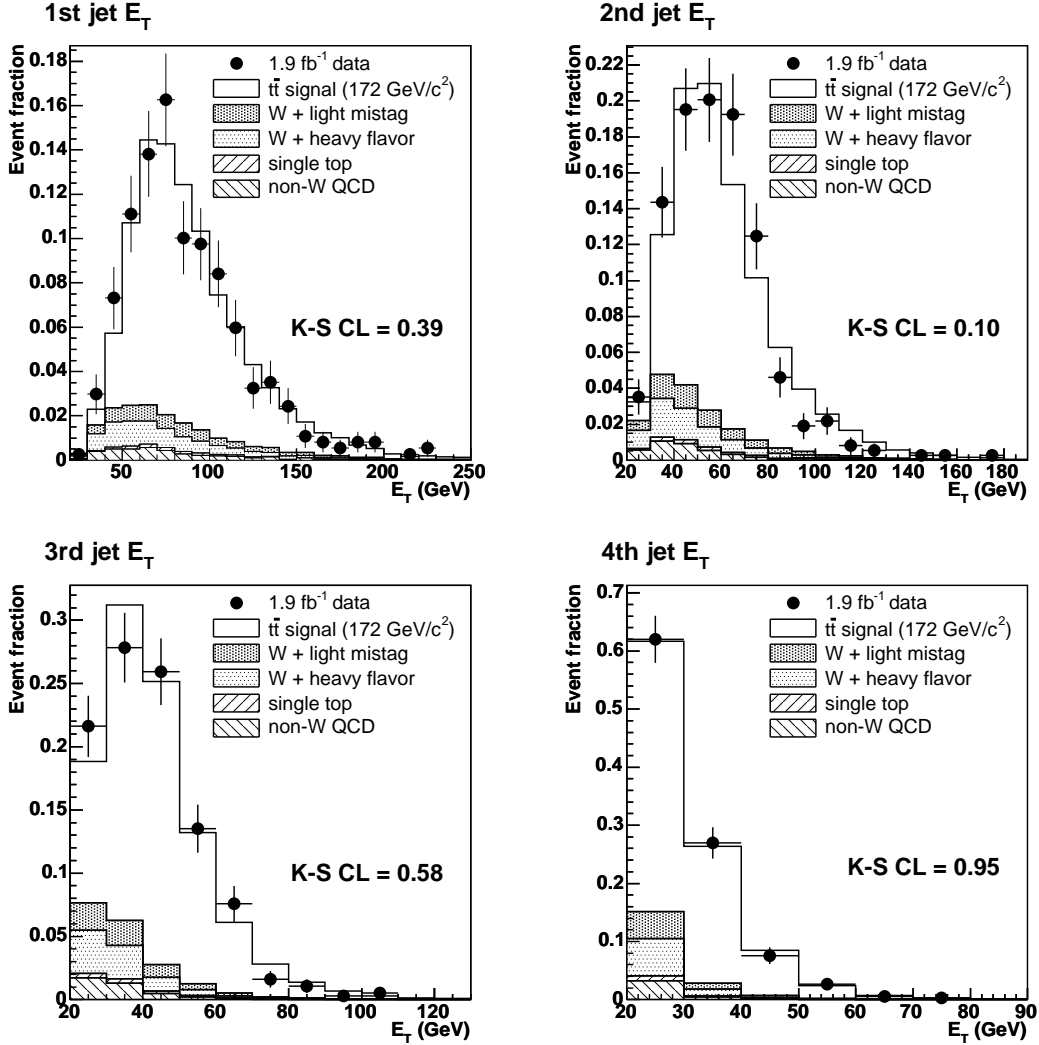}
  \caption{Comparison of data and MC predictions for the
  selected events. The confidence level obtained from a K-S test on
  the two distributions is indicated on the histogram. The plots show,
  in order, the corrected $E_T$ of the leading jet, 2nd jet, 3rd jet,
  and 4th jet in our events.}
\label{fig:val1-nocut}
\end{center}
\end{figure}

\begin{figure}[htbp]
\begin{center}
  \includegraphics*[width=0.9\textwidth,clip=true]{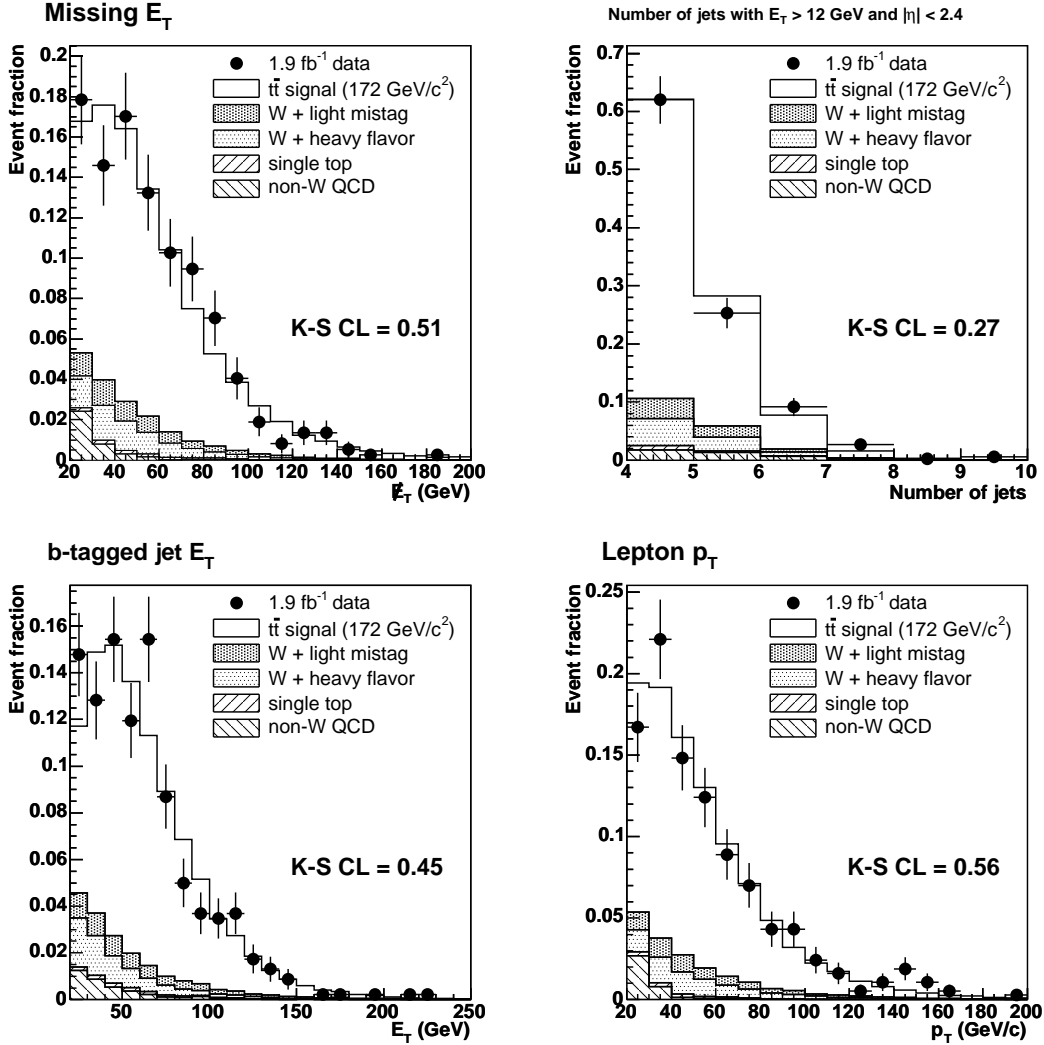}
  \caption{Comparison of data and MC predictions for the
  selected events. The confidence level obtained from a K-S test on
  the two distributions is indicated on the histogram. The plots show,
  in order, the event $\metnice$, the total number of jets with $E_T
  >$ 12 GeV and $|\eta| < 2.4$ in the event, the $E_T$ for all jets with
  a $b$-tag, and the lepton $p_T$.}
\label{fig:val2-nocut}
\end{center}
\end{figure}

Figures~\ref{fig:val1-nocut} and~\ref{fig:val2-nocut} show the
comparisons. All of these plots require exactly 4 jets with $E_T > 20$
GeV and $|\eta| < 2.0$, but in Fig.~\ref{fig:val2-nocut}, we also show
the number of jets with lower energies for an additional comparison
between data and MC. For each quantity, we perform a
Kolmogorov-Smirnov (K-S) test comparing the data and MC samples; in
all cases, the resulting confidence level shows a good agreement,
which validates the use of the MC generators and the QCD background
used in this analysis.

\section{Signal Likelihood}
\label{sec:analysis} 


The matrix element method allows for efficient incorporation of the
theoretical assumptions about the process under study into the data
analysis. The phase space integration procedure can be
viewed as a~Bayesian marginalization of the event probability over all
unobserved degrees of freedom. Particle theory provides a
well-motivated informative prior for this
marginalization. Maximization of the likelihood with respect to the
measured parameters results in an efficient (in the statistical sense)
estimate of these parameters.

Instead of attempting to integrate over the complete phase space of
the process which can include hundreds of particles, we assume that
the final state parton showering and hadronization processes, together
with the detector response, can be modeled empirically by the transfer
functions. This assumption allows for a drastic reduction in the phase
space complexity. However, calculation of the remaining tree-level
phase space integrals still remains a formidable problem when a large
number of events and parameter values must be processed. Therefore, we
employ additional assumptions, detailed later in this section, to
reduce the dimensionality. To compensate for these assumptions, we
introduce the concept of ``effective propagators'' that modify the
tree-level matrix element of the interaction.

\begin{figure}[htbp]
\centerline{
\includegraphics[width=.7\textwidth]{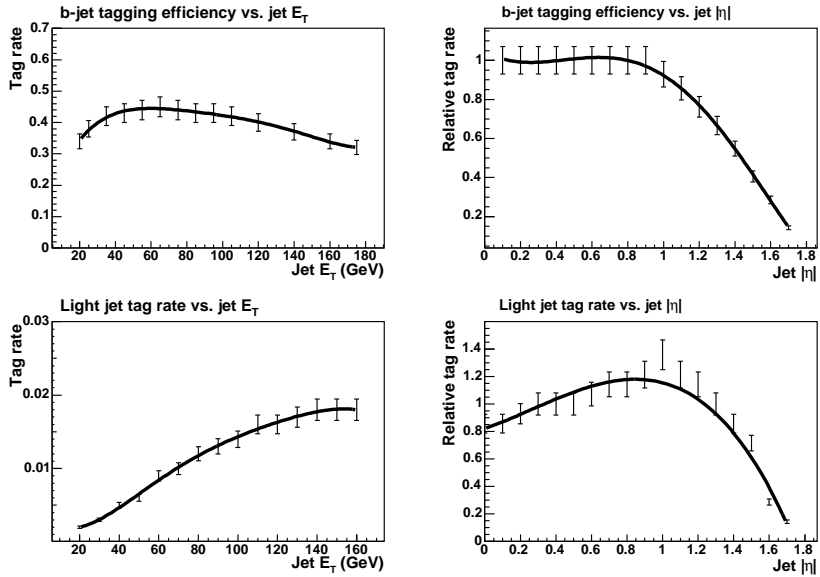}
}
\caption{Efficiency for the {\sc secvtx} algorithm with systematic uncertainties. The top two plots
show the tag efficiency for tagging $b$-jets as a function of the jet
$E_T$ (left) and $|\eta|$ (right), and the bottom two plots show the
tag rate for light jets (mistags), also as a function of jet $E_T$
(left) and $|\eta|$ (right). The fits used as a parameterization in
our analysis are also shown.}
\label{fig:secvtx_eff}
\end{figure}

For each event we obtain a $\ttbar$ signal probability as a function
of the top quark pole mass ($m_t$) and the jet energy scale (JES) using the
following expression:

\begin{multline}
  L(\vec{y} \mid m_t, \JES)
     = \frac{1}{N(m_t)} \frac{1}{A(m_t,\JES)} \times \\
  \sum_{i\,=1}^{24} w_{i} \int \frac {f(z_1) f(z_2)}{FF}
		      ~\textrm{TF}(\vec{y} \cdot \JES \mid \vec{x})
           ~|M_{\eff}(m_t,\vec{x})|^{2} ~d\Phi(\vec{x}),
\label{eq:prob_integral}
\end{multline}

\noindent
where $\vec{y}$ are the quantities we measure in the detector (the
momenta of the charged lepton and all the jets); $\vec{x}$ are the
parton-level quantities that define the kinematics of the event;
$N(m_t)$ is an overall normalization factor; ${A(m_t, \JES)}$ is the
event acceptance as a function of $m_t$ and JES; $f(z_1)$ and $f(z_2)$
are the parton distribution functions (PDF's) for incoming parton
momentum fractions $z_1$ and $z_2$; $FF$ is the relativistic invariant
flux; TF$(\vec{y} \cdot \JES
\mid \vec{x})$ are the transfer functions that predict the measured
jet momenta distributions given the quark kinematics; d$\Phi(\vec{x}$)
indicates integration over the phase space of the two initial and six
final state partons and leptons in the $\ttbar$ production and decay
(including necessary Jacobians); and $M_{\eff}$($m_t,\vec{x}$) is the
modified matrix element for $\ttbar$ production and decay. The PDFs,
$f(z_1)$ and $f(z_2)$, are integrated over the appropriate
combinations of incoming $q\bar{q}$ and gluons. We use the CTEQ5L
PDFs~\cite{cteq} in our integration. The integral is calculated for
each of the 24 possible permutations of jet-parton assignment and then
summed with the appropriate weights $w_i$, where the weights are
determined by the $b$-tagging information on the jets. Specifically,
for each tagged jet in the event, a weight equal to the tag rate of
the jet is given if it is assigned to a $b$ parton, and a weight equal
to the mistag rate is given if it is assigned to a light parton. An
untagged jet is given a weight of 1 minus the tag rate if assigned to
a $b$ parton, and 1 minus the mistag rate if assigned to a light
parton. The four individual jet weights are then
multiplied. Figure~\ref{fig:secvtx_eff} shows the parameterizations of
the tag rates as a function of jet $E_T$ and $|\eta|$ used to
determine the $w_i$ values; we assign a probability for a $c$ jet to
be tagged equal to 0.22 times the probability for a $b$-jet with the
same $E_T$ and $|\eta|$. These values are derived from Monte Carlo
events and then corrected with a scale factor measured in data to
account for differing tag rates in Monte Carlo events and data events.

We begin with the Kleiss-Stirling matrix element~\cite{kleiss}, which
includes both $q\bar{q} \rightarrow t\bar{t}$ and $gg \rightarrow
t\bar{t}$ production processes, as well as all spin correlations.  The
integral formula in Eq.~\ref{eq:prob_integral} requires a 24-dimensional 
integration
(eight four-vectors are needed to describe the reaction, but
energy-momentum conservation together with the negligible masses of the
initial partons and the final state leptons allow for a trivial
phase space dimensionality reduction to 24).
This is computationally difficult to evaluate, so we make the
following simplifying assumptions: the lepton direction and momentum
are perfectly measured; the directions of the partons coincide with
the measured jet directions; the light quark masses are zero, the $b$
quark from the hadronic top quark is on mass shell, and the $b$ quark
from the leptonic top quark has zero mass. The last assumption results
in a simplification of the kinematic equation on the leptonic side,
from an 8th-order to a 4th-order polynomial. We introduce a prior for
the transverse momentum of the $\ttbar$ system into the overall event
probability formula, but we do not consider the transverse motion of
individual initial partons.  This allows us to eliminate two more
integration variables. The $\ttbar$ transverse momentum prior, as
constructed from {\sc herwig} MC samples, is nearly independent of the
top quark mass for masses between 120 and 220 GeV/$c^2$. The
transverse momentum of the initial partons is also neglected for the
determination of $z_1$ and $z_2$ from the event kinematics.

We choose the set of seven remaining variables of integration to be
the squared masses $M_t^2$ and $M_W^2$ on both the leptonic decay and
hadronic decay side of the $t \bar t$ system, $\beta =\log(p_q /
p_{\bar q})$, where $p_q$ and $p_{\bar q}$ are the magnitudes of the momenta of the
two products from the hadronic $W$ boson decay, and the
two-dimensional transverse momentum vector $\vec{p}_T$ of the $\ttbar$
system. Note that the top quark and $W$ boson pole masses, $m_t$ and
$m_W$, are not the same variables as $M_t$ and $M_W$. The latter
variables refer to the top quark and $W$ boson masses in a given
event, and we integrate over them in our likelihood calculation.

The expected distributions of $M_t^2$ and $M_W^2$ are defined almost
exclusively by the top quark and $W$ boson propagator terms in the
matrix element. Nominally, these are relativistic Breit-Wigners peaked
at the top quark and $W$ boson masses; however, the kinematic
assumptions to reduce the number of integration dimensions described
above cause the $M_t^2$ and $M_W^2$ distributions to be altered from
their Breit-Wigner form. To account for this, we replace the
Breit-Wigner propagators in the matrix element with propagators that
reflect the assumptions; we call these adjusted propagators
``effective propagators.''

\subsection{Effective Propagators}
\label{sec:effec-prop}

The effective propagators are built by calculating invariant masses of
``effective partons.'' These objects are constructed in such a manner
that their four-momenta can be reproduced exactly using only the
integration variables and the variables measured in the detector by
solving kinematic equations consistent with our assumptions.
In each MC event, we find the assignment
of the four tree-level partons to the four highest $p_T$ jets reconstructed
in the detector that minimizes the combined distance in the $\eta$--$\phi$ space
between partons and jets. Then we
construct effective partons by building four-vectors that have
the energies of the tree-level partons, the directions of
the matched calorimeter jets, and the masses used in the kinematic equation
solvers ({\it i.e.}, zero for light quarks and the leptonic side $b$, and
4.95~GeV/$c^2$ for the hadronic side $b$). We associate effective values of
top quark and $W$ boson mass, as well as $\beta$ and
$\vec{p}_{T}(t \bar t$\,), by building
these quantities out of the effective partons.
The effective partons are also used in the construction of our
calorimeter transfer functions, for the sake of consistency.

The construction of the effective propagator on the leptonic side uses
values of the lepton momentum smeared according to the Gaussian
resolution functions given in Section~\ref{sec:detect}:
$\sigma(p_T)/p_T = 0.1\% \cdot p_T/(\mathrm{GeV}/c)$ for muons and
$\sigma(E_T)/E_T$ = 13.5\%/$\sqrt{E_T/\mathrm{GeV}}~\oplus$ 2\% for
electrons.

The choice of the neutrino $p_{z}$ is ambiguous; for each event we
solve for the $p_{z}$ that minimizes the deviation of the effective
leptonic $W$ boson and top quark masses, $M_{W,\textrm{eff}}$ and
$M_{t,\textrm{eff}}$, from the tree-level $W$ boson mass, $M_{W,\textrm{gen}}$,
and top quark mass, $M_{t,\textrm{gen}}$.  The deviation is quantified by a
$\chi^{2}$ defined as
\begin{eqnarray}
\chi^{2} = \frac{(M_{t,\mathrm{eff}}^2 - M_{t,\mathrm{gen}}^2)^2}{\sigma_{t}^2} + \frac{(M_{W,\mathrm{eff}}^2 - M_{W,\mathrm{gen}}^2)^2}{\sigma_{W}^2},
\end{eqnarray}
where
\begin{eqnarray}
\sigma_{t} = \Gamma_{t} m_{t}, \ \ \ \ \sigma_{W} = \Gamma_{W} m_{W},
\end{eqnarray}
$m_{t}$ and $m_{W}$ are pole masses for the top quark and the
$W$ boson, and $\Gamma_t$ and $\Gamma_W$ are their decay widths.
When this $p_{z}$ search is performed, the transverse momentum of the
leptonic-side top quark is set to the difference between 
the tree-level MC value of the $t\bar t$
transverse momentum and the effective transverse
momentum of the hadronic-side top quark.

In our calculations we assume that there is no correlation between the
effective propagators on the hadronic and leptonic sides of the
event. In reality, the invariant masses on the leptonic side are
affected by the hadronic side uncertainties due to the definition of
the leptonic top quark momentum used in the effective propagator
construction. However, the uncertainty due to the transverse momentum
transfer from the hadronic to the leptonic side is not large in
comparison with the uncertainty associated with the unknown transverse
momentum of the $t\bar t$ system itself.

The widening of the effective invariant masses in comparison with the
original Breit-Wigner distributions depends on the event kinematics.
For example, finite angular resolution of the detector results in
the greatest widening of the effective $W$ boson mass distribution on
the hadronic side when the opening angle between the two jets
originating from the $W$ boson decay is close to $\pi/2$.
Due to the large dimensionality of the phase space and limited CPU
resources available, it is not feasible to model the shapes of the
effective invariant mass distributions for each kinematic
configuration encountered in the calculation of event probabilities.
Instead, we characterize these shapes using a low-dimensional quantity
with high predictive power: the covariance matrix for effective $W$
boson and top quark masses.

\begin{figure}[htbp]
\begin{center}
\epsfig{file=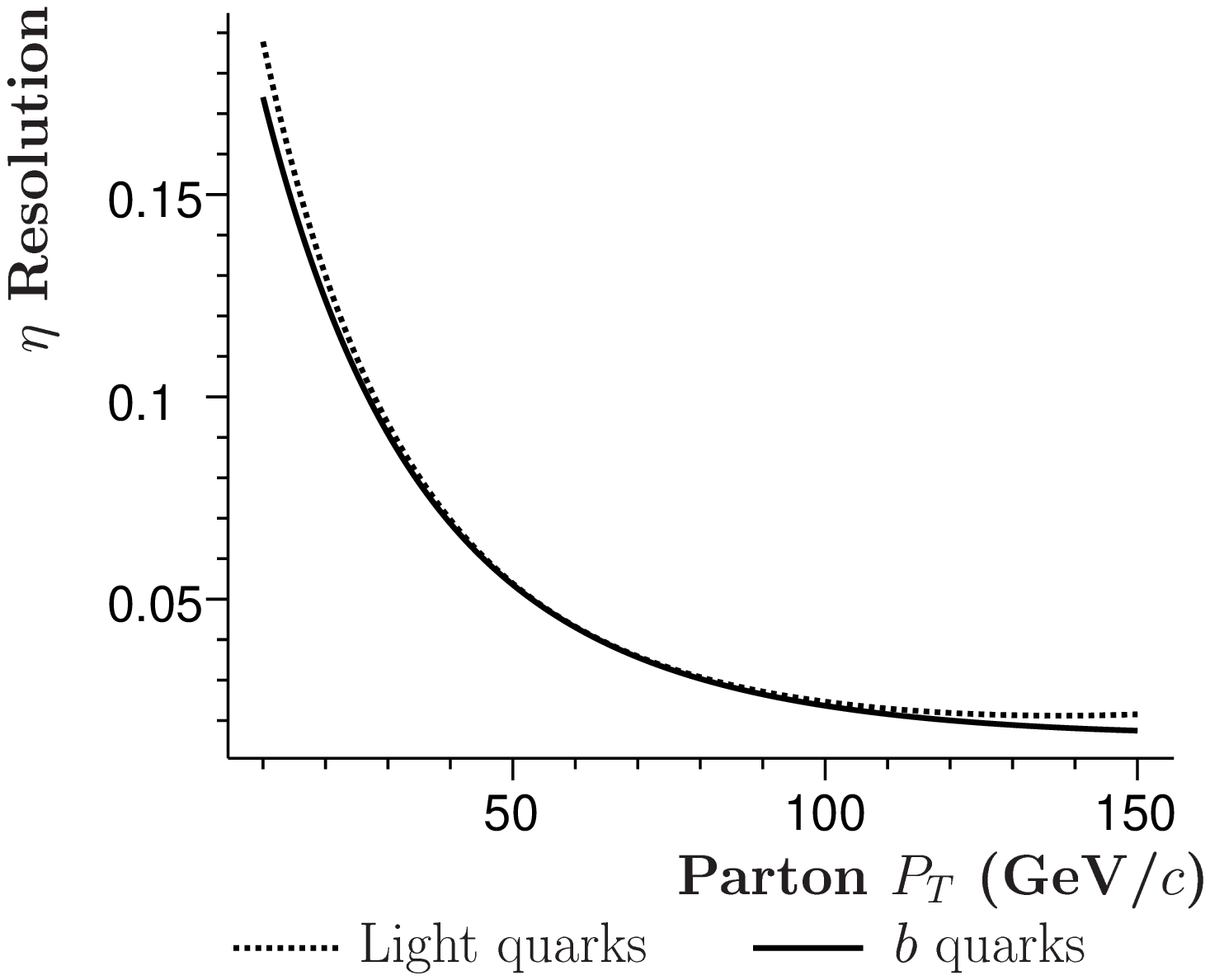,width=0.4\textwidth}
\epsfig{file=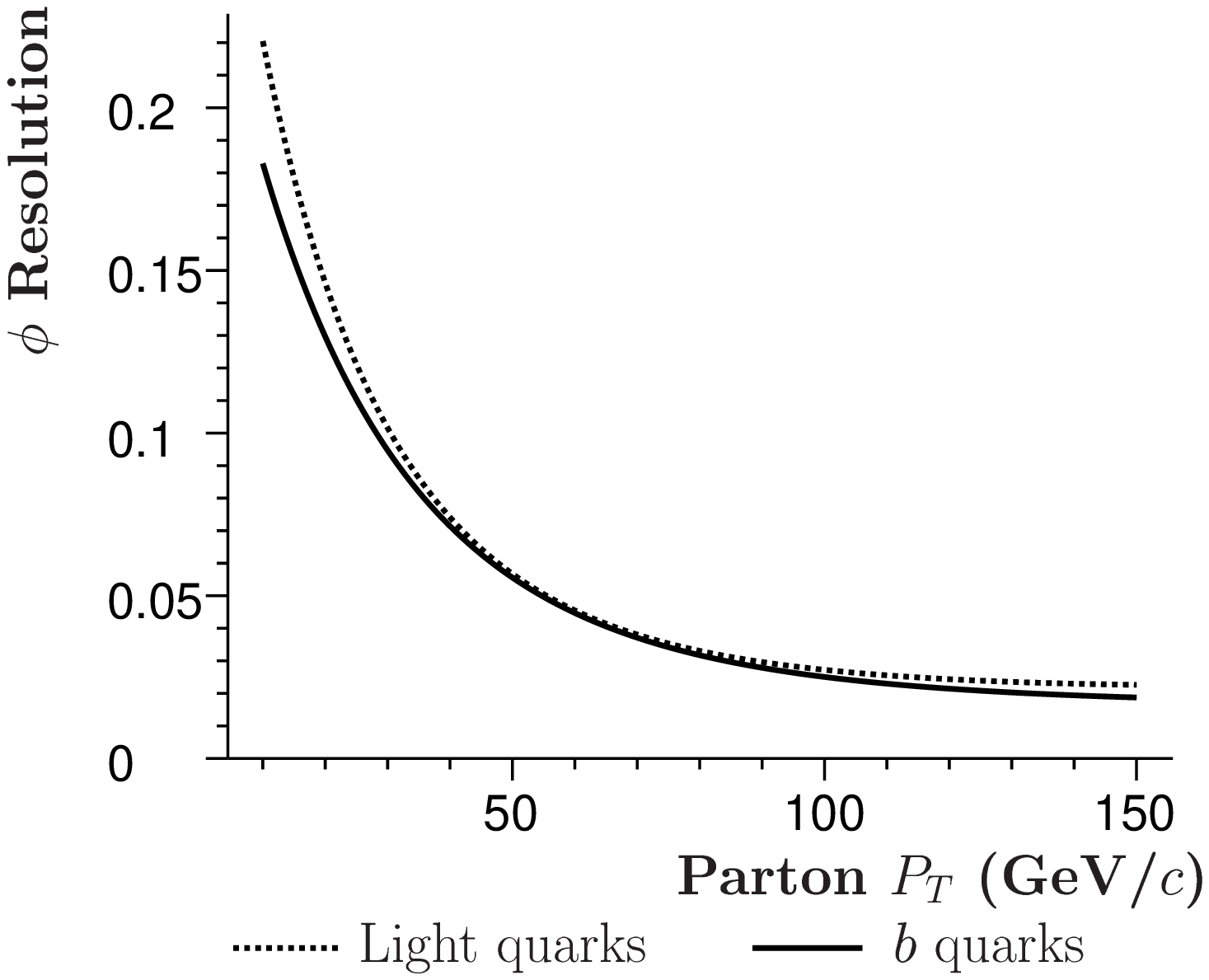,width=0.4\textwidth}
\caption{Angular resolution in $\eta$ (left) and $\phi$ (right) as a
function of parton $p_T$. The dashed line indicates the resolution for
light quarks, and the solid line for $b$ quarks.}
\label{fig:angular_resolution}
\end{center}
\end{figure}

On the hadronic side of the event we calculate the appropriate
Jacobian and propagate the uncertainties from jet masses and angles to
$M_{t}^{2}$ and $M_{W}^2$ using the standard, first-order multivariate
error propagation formulae. We assume that the uncertainties of jet
masses and angles are not correlated.  The angular resolutions
calculated with {\sc herwig} and used to build the hadronic-side
propagators are shown in Fig.~\ref{fig:angular_resolution}, where
the angular resolution is defined as the width of the $\Delta \eta$ or
$\Delta \phi$ distribution in a given parton $p_T$ bin. The resolution
on the jet mass squared used for covariance matrix estimation is
assumed to be constant: $\sigma (m^2) = 242$ GeV$^2/c^4$ for $b$-jets
and $\sigma (m^2) = 202$ GeV$^2/c^4$ for light jets.  (These values
are also calculated with {\sc herwig}.) The uncertainties on the
magnitudes of the parton momenta are not used to build this covariance
matrix --- these uncertainties are taken into account by the transfer
functions.

Three independent quantities can be extracted from the covariance
matrix constructed in this manner: the standard deviations for
$M_{t}^{2}$ and $M_{W}^2$ and the correlation coefficient. We assume
that the hadronic side effective propagator depends only on these
three quantities. However, we do not make any functional assumption
about the propagator shape. Instead, we use MC events to build a
non-parametric estimate of the propagator density. We split the 3-D
space of the two standard deviations and the correlation coefficient
into cells that contain approximately equal numbers of events.  In
each cell, we use a kernel density estimation technique~\cite{ref:kde}
to construct the effective propagator. The propagator is initially
evaluated on a sufficiently dense rectangular grid, and fast linear
interpolation is used to find its values during subsequent
calculations.  When the event is reconstructed during the likelihood
integration, the covariance matrix is calculated for the given
kinematics, and the corresponding effective propagator density is
looked up in a table of pre-computed propagators.

\begin{figure}[htbp]
\centerline{
\epsfig{file=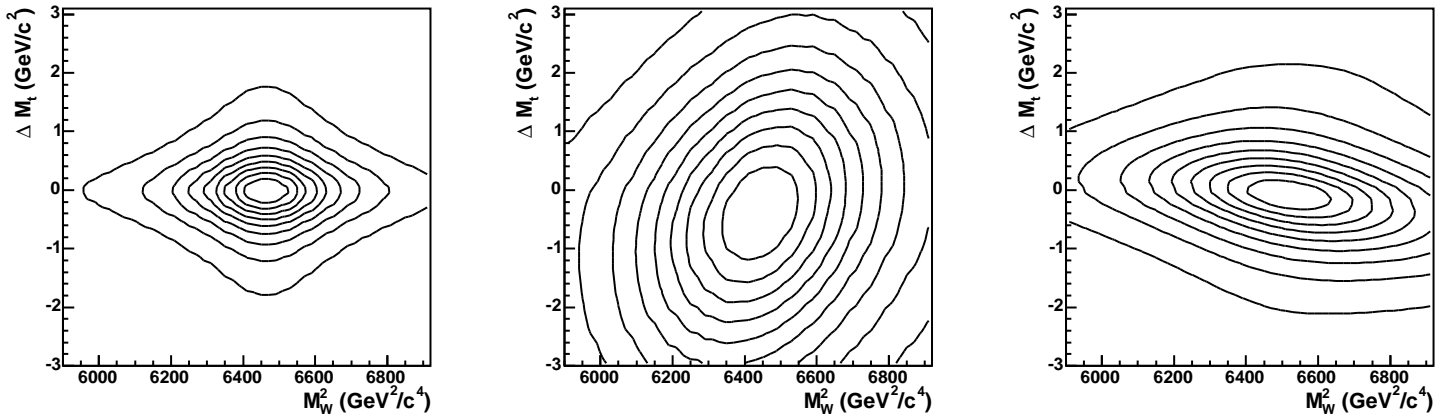}
}
\caption{Contour plots of the propagators in $M_W^2$ and $\Delta
M_t$. Left: Breit-Wigner propagator, for comparison. Center: Sample
effective propagator on the hadronic side of the event. Right: Sample
effective propagator on the leptonic side of the event.}
\label{fig:eff_propagators}
\end{figure}

Due to the presence of unobserved neutrino, the kinematic
configuration on the leptonic side of the event is significantly less
constrained than that on the hadronic side, and provides less information
about the top quark mass and the jet energy scale. Because of this, we
employ a~simplified model for the leptonic side effective propagators.
These propagators are averaged over various kinematic configurations,
and they depend only on the assumed top quark pole mass, not on the
kinematics of a particular event. (The hadronic side propagators
depend on the top quark pole mass implicitly, via the Jacobians used
in the error propagation.)  Examples of the hadronic and leptonic side
effective propagators are shown in Fig.~\ref{fig:eff_propagators}.

\subsection{Transfer Functions}
\label{sec:transfer}
\begin{figure}
\epsfig{file=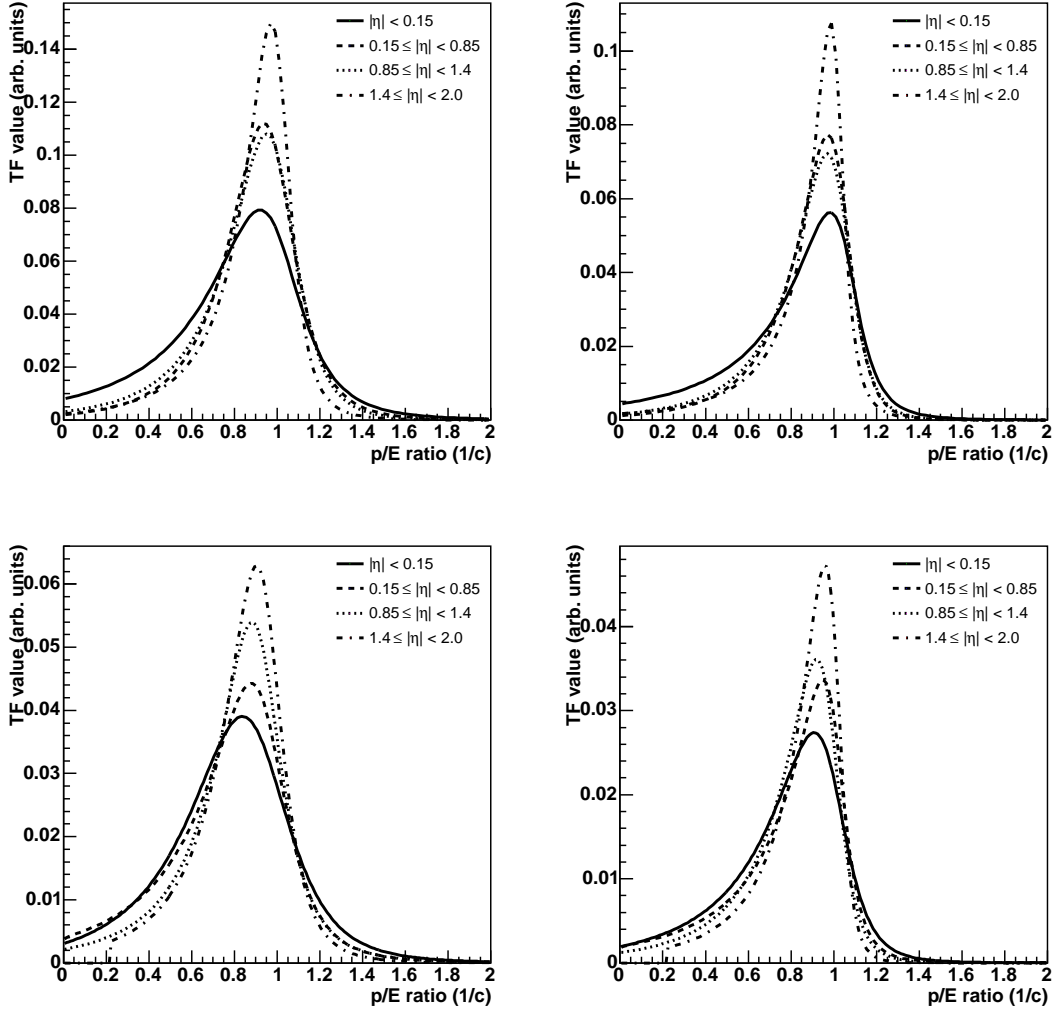, width=.9\textwidth}
\caption{Sample fitted transfer functions for light and $b$
  quarks in $\eta$ bins.  Transfer functions are shown for light
  quarks with parton $P_T=40$ and 70 GeV/$c$ (top left and right,
  respectively), and for $b$ quarks with parton $P_T=40$ and 70
  GeV/$c$ (bottom left and right, respectively).}
\label{fig:tf_sample}
\end{figure}

The transfer functions relate the parton transverse momentum, $p_T$, to 
the measured jet momentum.
They
are probability distributions of $p$/$E$, the ratio of the magnitude 
of the jet
momentum $p$ to the parent effective parton energy $E$, parametrized
as a function of the transverse momentum of the parent effective
parton (described in the previous section). We construct our transfer
functions using $\ttbar \rightarrow$ lepton + jets MC events in
a wide range of top quark masses, requiring the same selection cuts as
described earlier. In this sample, the parton is  matched to the
simulated jets, $p/E$ distributions
are created in bins of the parent parton $p_T$, and then these
distributions are fit with a $p_T$-parametrized function. The
function is constructed using Johnson curves~\cite{johnsf}, which
allow us to fit a variety of non-Gaussian shapes.
These curves are parametrized by quantities
calculated from the transfer function distributions themselves:
mean ($\mu$), standard deviation ($\sigma$), skewness ($s$), and
kurtosis ($k$), which, in turn, smoothly depend on the parton $p_T$.
We extrapolate the fitted transfer
functions for momenta that
are below the cutoff value imposed in the sample. This extrapolation
ensures that the transfer functions are correctly normalized, as it
accounts for jets that do not appear in our sample due to selection
cuts. Separate transfer functions are created for four different
$\eta$ regions of the detector, as well as for $b$ quarks and light
quarks. Figure~\ref{fig:tf_sample} shows transfer function examples
for light and $b$ quarks for several different parton $p_T$ values
and $\eta$ bins. \\

\subsection{Normalization and acceptance}
\label{norm-accept}

\begin{figure}[htbp]
\begin{center}
\epsfig{file=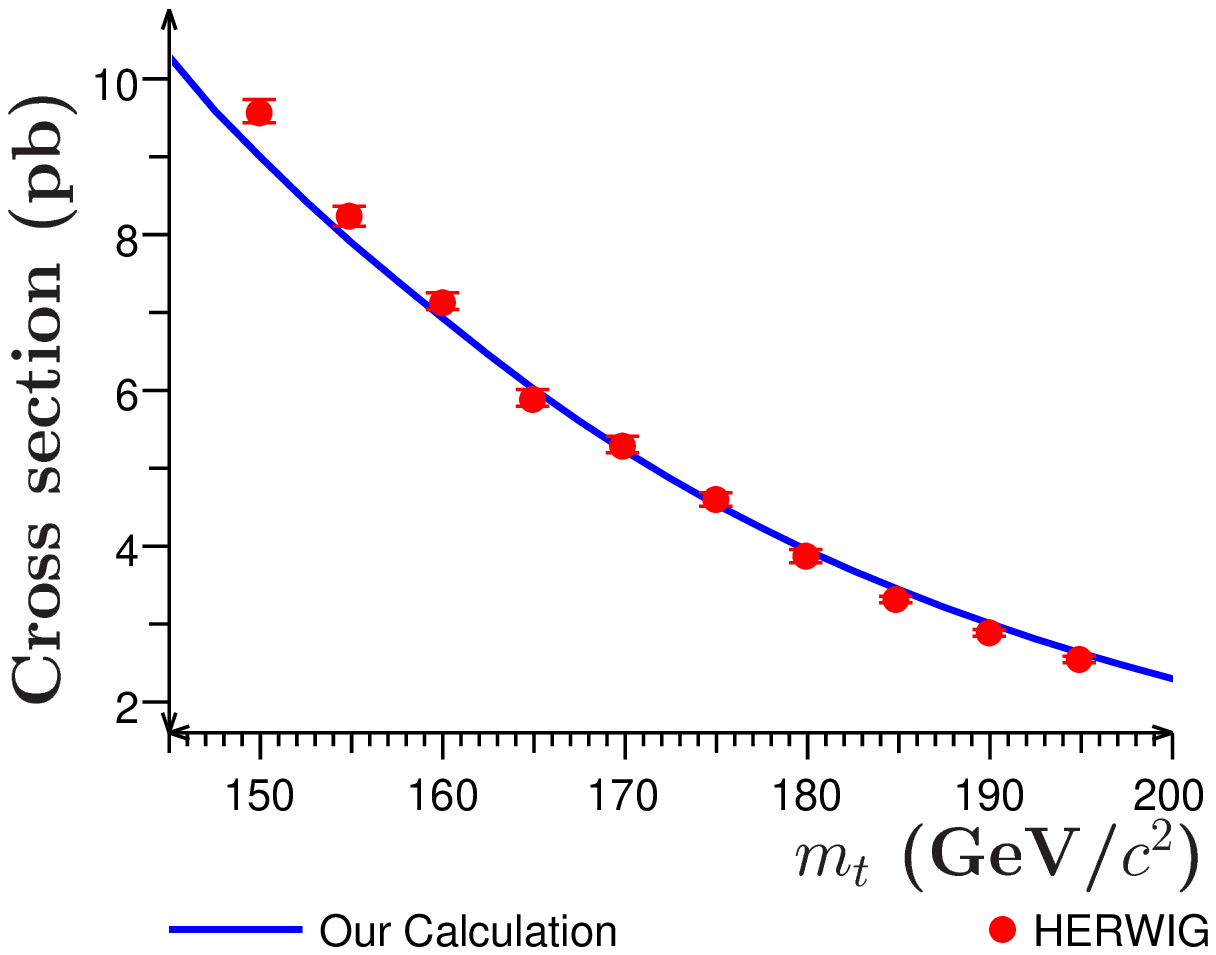,width=0.45\textwidth}
\epsfig{file=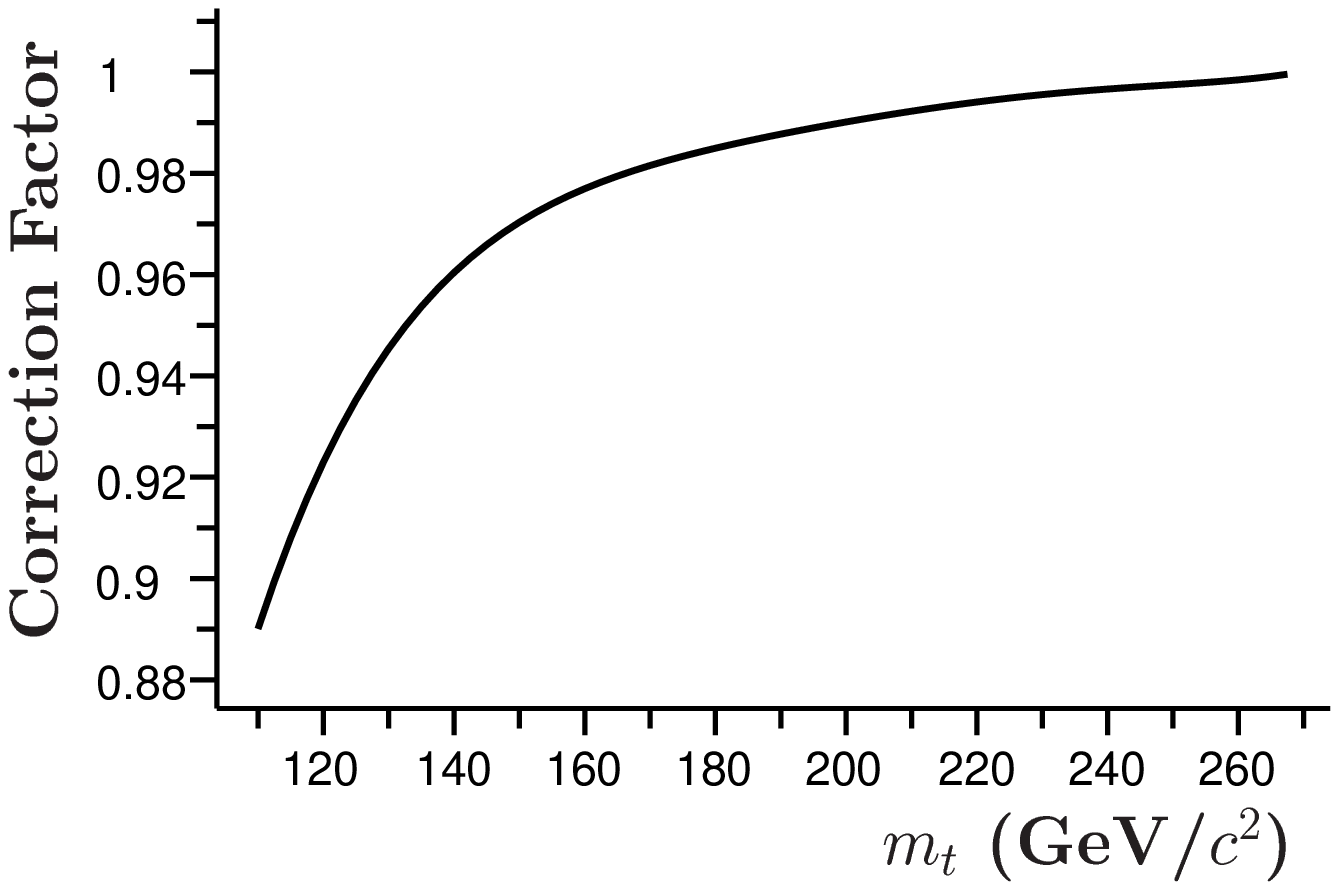,width=0.45\textwidth}
\caption{\label{fig:norm} Left: Cross section obtained from our
normalization calculation as a function of $m_t$ compared with the
cross section used in {\sc herwig}. Right: Normalization correction
factor due to effective propagators as a function of $m_t$.}
\end{center}
\end{figure}

The normalization factor $N(m_t)$ in Eq.~\ref{eq:prob_integral}
is obtained by integrating the Kleiss-Stirling matrix
element~\cite{kleiss} together with the PDF's and the flux factor over
the phase space formed by the two initial and the six final state
particles. The resulting cross section as compared to the $\ttbar$
cross section in {\sc herwig} is shown in Fig.~\ref{fig:norm}. We
do not expect perfect agreement due to the absence of radiation in our
code.

The resulting normalization is then corrected by a~small
additional factor to account for the difference between the effective
propagators and Breit-Wigners.  The normalization correction is
calculated as follows. We split the matrix element into two parts:
$|M(\vec{x})|^2 = \Gamma_{\tree}(\vec{x}) |M(\vec{x})^*|^2$, where
$\vec{x}$ completely specifies the phase space point,
$\Gamma_{\tree}(\vec{x})$ is the product of the four Breit-Wigners,
and $|M(\vec{x})^*|^2$ is the rest of the matrix element. Using this
notation, the tree-level normalization is $\int F(\vec{x})
\Gamma_{\tree}(\vec{x}) |M(\vec{x})^*|^2 d\vec{x}$, where $F(\vec{x})$
is the remaining term in the cross section (flux factor and structure
functions). The correct normalization with the effective propagators
is instead

\begin{eqnarray}
\int F(\vec{x}_{\eff}) \Gamma_{\eff}(\vec{x}_{\eff}) |M(\vec{x}_{\eff})^*|^2 d\vec{x}_{\eff},
\end{eqnarray}

\noindent where $\vec{x}_{\eff}$ specifies the kinematic configuration of 
the effective partons. This quantity can be rewritten as 

\begin{eqnarray}
\int \frac{F(\vec{x}_{\eff}) \Gamma_{\eff}(\vec{x}_{\eff}) |M(\vec{x}_{\eff})^*|^2 d\vec{x}_{\eff}}{F(\vec{x}) \Gamma_{\tree}(\vec{x}) |M(\vec{x})^*|^2 d\vec{x}} F(\vec{x}) \Gamma_{\tree}(\vec{x}) |M(\vec{x})^*|^2 d\vec{x}. 
\end{eqnarray}

\noindent
From this point forward we proceed as if there is a one-to-one
correspondence between points in $\vec{x}$ and
$\vec{x}_{\eff}$~\footnote{This assumption is not true in general.
The $\vec{x} \rightarrow \vec{x}_{\eff}$ mapping involves smearing due to
hadronization and angular resolution, so the mapping is many-to-many.
Our assumption produces just a first-order correction. However,
the correction appears to be small, so we have stopped at this level.}.
By construction, the propagators play the role of densities
in their corresponding
spaces. Therefore, 
$\Gamma_{\eff}(\vec{x}_{\eff}) d\vec{x}_{\eff} = \Gamma_{\tree}(\vec{x}) d\vec{x}$, and we need only to calculate

\begin{eqnarray}
 \int \frac{F(\vec{x}_{\eff}) |M(\vec{x}_{\eff})^*|^2}{F(\vec{x}) |M(\vec{x})^*|^2} F(\vec{x}) \Gamma_{\tree}(\vec{x}) |M(\vec{x})^*|^2 d\vec{x}.
\end{eqnarray}

\noindent
This quantity is the average value of 

\begin{eqnarray}
\frac{F(\vec{x}_{\eff}) |M(\vec{x}_{\eff})^*|^2}{F(\vec{x}) |M(\vec{x})^*|^2},
\end{eqnarray}

\noindent
calculated over the tree-level MC events, times the cross section.
The resulting correction factor is plotted in Fig.~\ref{fig:norm} as a
function of the top quark mass.

\begin{figure}[ht]
\begin{center}
\includegraphics*[width=.7\textwidth]{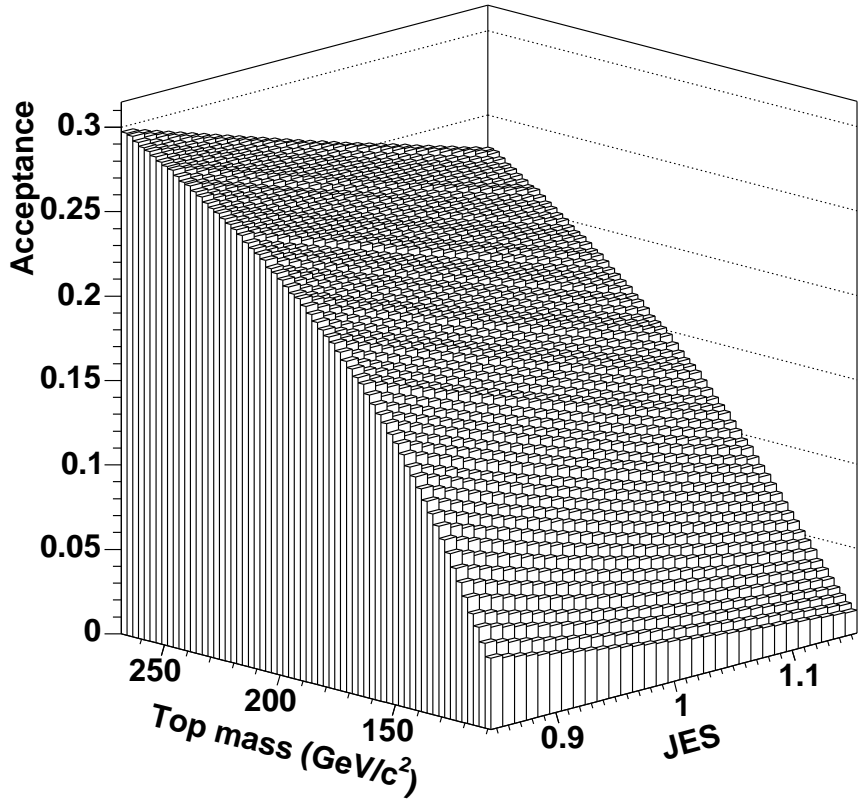}
\caption{Acceptance used in the integration as a function of $m_t$ and
JES.}
\label{fig:efficiency_latest}
\end{center}
\end{figure}

The acceptance $A(m_t, \JES)$ is obtained from $\ttbar$ MC events in
which parton angles are randomized to simulate the small angular resolution
uncertainty of the detector, and parton momenta are smeared
according to our transfer functions to mimic the jet
momenta that would have been measured in the detector. The
kinematic distributions for the smeared events are
similar to those of fully simulated events.
We do this for all the values of the top quark
mass and JES over which the likelihood function is defined, and then
calculate the acceptance at each $m_t$ and JES value to be the
fraction of these MC events that pass our selection cuts.
The advantage of this approach as opposed to using fully simulated
MC events is that the jet-parton association is exact, and events
with incorrect jet-parton association can be excluded from the
efficiency calculations. Our probability model describes tree-level
signal events with the correct set of jets; therefore, we do not use
fully simulated events, which include effects not accounted for in our
model, such as gluon radiation. 
The transfer functions are normalized with respect to all jet
momenta, not just those which pass the cuts. By building an acceptance
function from events smeared according to our transfer functions,
we directly normalize our likelihood.
Furthermore, we can generate our acceptance
from a much larger sample of events because we avoid the computing intensive
steps of event simulation and reconstruction, while reducing statistical
fluctuations in the resulting curve.
Figure~\ref{fig:efficiency_latest} shows the 2-D
acceptance as a function of $m_t$ and JES. 

\section{Background Discrimination and Final Likelihood}
\label{sec:tot-likeli}

Our integration method calculates the likelihood for an event assuming
the $\ttbar$ hypothesis, hence we use a neural network approach to
discriminate against background events.  The neural network uses ten
inputs: seven variables describing the kinematics of the event (the
$p_T$ for the four leading jets, the lepton $E_T$, $\metnice$, and
$H_T$, the scalar sum of these quantities), and three variables
describing the topology of the event.  The three topological variables
are aplanarity, $D_R$, and $H_{\textrm{TZ}}$. The aplanarity is
defined as $3/2 Q_1$, where $Q_1$ is the smallest eigenvalue of the
normalized momentum tensor $\Theta_{ab} =
\sum_i p^i_a p^i_b / \sum_i |\vec{p_i} |^2$, where the indices $a$ and $b$ run
over the three axes $x$, $y$, and $z$, the index $i$ runs over the
four jets and charged lepton, and $\vec{p_i}$ is the three-momentum of
a given particle. The variable $D_R = \Delta R_{ij}^{\textrm{min}}
\cdot
\min(p_z^{(i,j)}) / p_T^{\ell}$, where $\Delta R_{ij}^{\textrm{min}}$
is the smallest $\Delta R$ between any pair of jets,
$\min(p_z^{(i,j)})$ is the smaller of the two $p_z$ values for the two
jets in that pair, and $p_T^{\ell}$ is the transverse momentum of the
charged lepton.  $H_{\textrm{TZ}}$ is a ratio of scalar sums of
transverse and longitudinal momenta; the numerator contains all jets
except the leading jet, and the denominator sums all jets, the charged
lepton, and the neutrino. The smaller $|p_z^{\nu}|$ solution given by
the kinematic equation for the leptonic $W$ boson decay (assuming $M_W
= 80.4$ GeV/$c^2$) is taken.  The ten variables are summarized in
Table~\ref{NN_var}. To construct the neural network, we use the
{\sc jetnet} neural network package, version 3.5~\cite{jetnet}.

\begin{table} [htbp]
\caption{Variables included in the neural network discriminant: the first seven
are kinematic variables, the last three are topological variables.}
\label{NN_var}
\begin{center}
\begin{tabular}{ll}
\hline \hline
Variable      & Definition  \\  \hline
$p_T^i$       & $p_T$ of each of the 4 leading jets        \\
$E_T^{\ell}$   & Charged lepton $E_T$ (electron) or $p_T$ (muon) \\
$\metnice$    & The missing $E_T$ \\
$H_T$     & Scalar sum of jets and lepton transverse momenta and $\metnice$ \\
Aplanarity $= 3/2 \cdot Q_1$ & $Q_1$: smallest eigenvalue of the momentum tensor\\
$D_R = \Delta R_{ij}^{\textrm{min}} \cdot \min(p_z^{(i,j)}) / p_T^{\ell}$ &
        $\Delta R_{ij}^{\textrm{min}}$ is the smallest $\Delta R$ between any pair of jets \\
$H_{\textrm{TZ}}= \sum_{i=2}^{4} |p_T^i| / (\sum_{i=1}^{4} |p_z^i| +
|p_z^{\ell}| + |p_z^{\nu}|)$ & Ratio of scalar sums of transverse to longitudinal momenta \\
\hline \hline
\end{tabular}
\end{center}
\end{table}

\begin{figure}
\centerline{
\epsfig{file=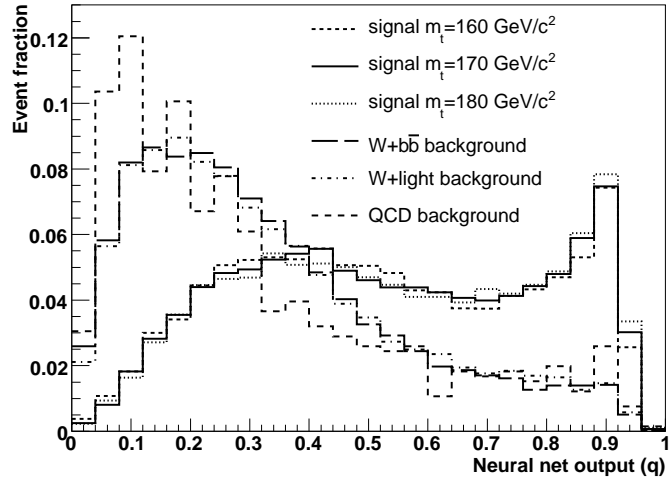,width=.6\textwidth}
}
\caption{\label{fig:nn_output} The distributions of our neural network
  discriminant variable for signal and background MC events. The peaks
  on the right are from $\ttbar$ signal events at three top masses. 
  The peaks on the left are from various types of backgrounds.}
\end{figure}

The neural network is trained to separate $\ttbar$ events with a mass
of 170 GeV/$c^2$ from $W + b\bar{b}$ background; we then cross-check
the neural network with other signal masses and background types to
make sure that the output shape is not dependent on the signal mass
present. Figure~\ref{fig:nn_output} shows the neural network output,
$q$, for a variety of different samples.  We compute the background
fraction for each observed event as $f_{\bg}(q) = B(q)/[B(q) + S(q)]$,
where the background distribution $B(q)$, obtained by adding each type
of background with its own weight, and signal distribution $S(q)$ are
each normalized to their overall expected fractions.

Our total likelihood from all events will naturally contain
likelihoods from signal events and background events. However, only
the signal events will contain meaningful information about
$m_t$. Thus, we want to remove the contribution due to background
events from the total likelihood to recover the likelihood from signal
events. (Note that there is not a separate matrix element for
background processes --- the likelihood for all events, signal and
background, is calculated under the assumption that the event is a
$\ttbar$ signal event.) Consequently, we compute from Monte Carlo
simulation the average likelihood for background events and subtract
out the expected contribution due to background events from the total
likelihood:

\begin{eqnarray}
 \log L_{\textrm{mod}}(m_t, \JES) = \sum_{i \in \textrm{events}}
[\log L(\vec{y_i} | m_t, \JES)] -
n_{\bg} \log
\overline{L_{\textrm{bg}}}(m_t, \JES),
\label{eq:bg_subtraction_n}
\end{eqnarray}

\noindent
where $L_{\textrm{mod}}$ is the modified total likelihood for a given
set of events, $L(\vec{y_i} | m_t, \JES)$ is the likelihood for an
individual event, $n_{\bg}$ the expected number of background events,
and $\overline{L_{\textrm{bg}}}(m_t, \JES)$ is the average likelihood
for a background event as computed in Monte Carlo simulation. This
calculation is performed separately for 1-tag and $>$1-tag events, as
the background fractions and $\overline{L_{\textrm{bg}}}(m_t, \JES)$
are different for the two subsamples.

We can rewrite Eq.~\ref{eq:bg_subtraction_n} in terms of the individual
per-event background fraction to obtain our final modified likelihood
$L_{\textrm{mod}}$:

\begin{eqnarray}
\log L_{\textrm{mod}}(m_t, \JES) = \sum_{i \in \textrm{events}}
  [\log L(\vec{y_i} | m_t, \JES) - f_{bg}(q_i) \log
  \overline{L_{\textrm{bg}}}(m_t, \JES)],
\label{eq:final_bg_subtraction}
\end{eqnarray}

\noindent
where $f_{\bg}(q_i)$ is the background fraction given the discriminant
variable $q_i$ for a given event. Equations~\ref{eq:bg_subtraction_n}
and~\ref{eq:final_bg_subtraction} are equivalent if the number of
background events in the data is equal to the expected background
contribution. However, the advantage of using
Eq.~\ref{eq:final_bg_subtraction} is that if there are more or
fewer background-like events in our data than expected, the average
value of $f_{\bg}(q_i)$ will be correspondingly higher or lower, thus
compensating for the difference.

There is another class of events not well-modeled by our signal
likelihood integration or handled by the background subtraction above,
which we call ``bad signal'' events. These are $\ttbar$ signal events
in which the four observed jets and/or lepton are not directly
produced from the $t\bar{t}$ decay. These events exist due to a variety
of causes (extra jets from gluon radiation, $\ttbar$ events where both
$W$ bosons decay leptonically or hadronically, $W \rightarrow
\tau\nu$ decay, etc.) and comprise roughly 35\% of our total
signal. For a signal mass of 172 GeV/$c^2$, 36.2\% of the single-tag
and 30.9\% of the $>$1-tag events fall into the ``bad signal''
category.  

\begin{figure}
\begin{center}
    \includegraphics*[width=.45\textwidth]{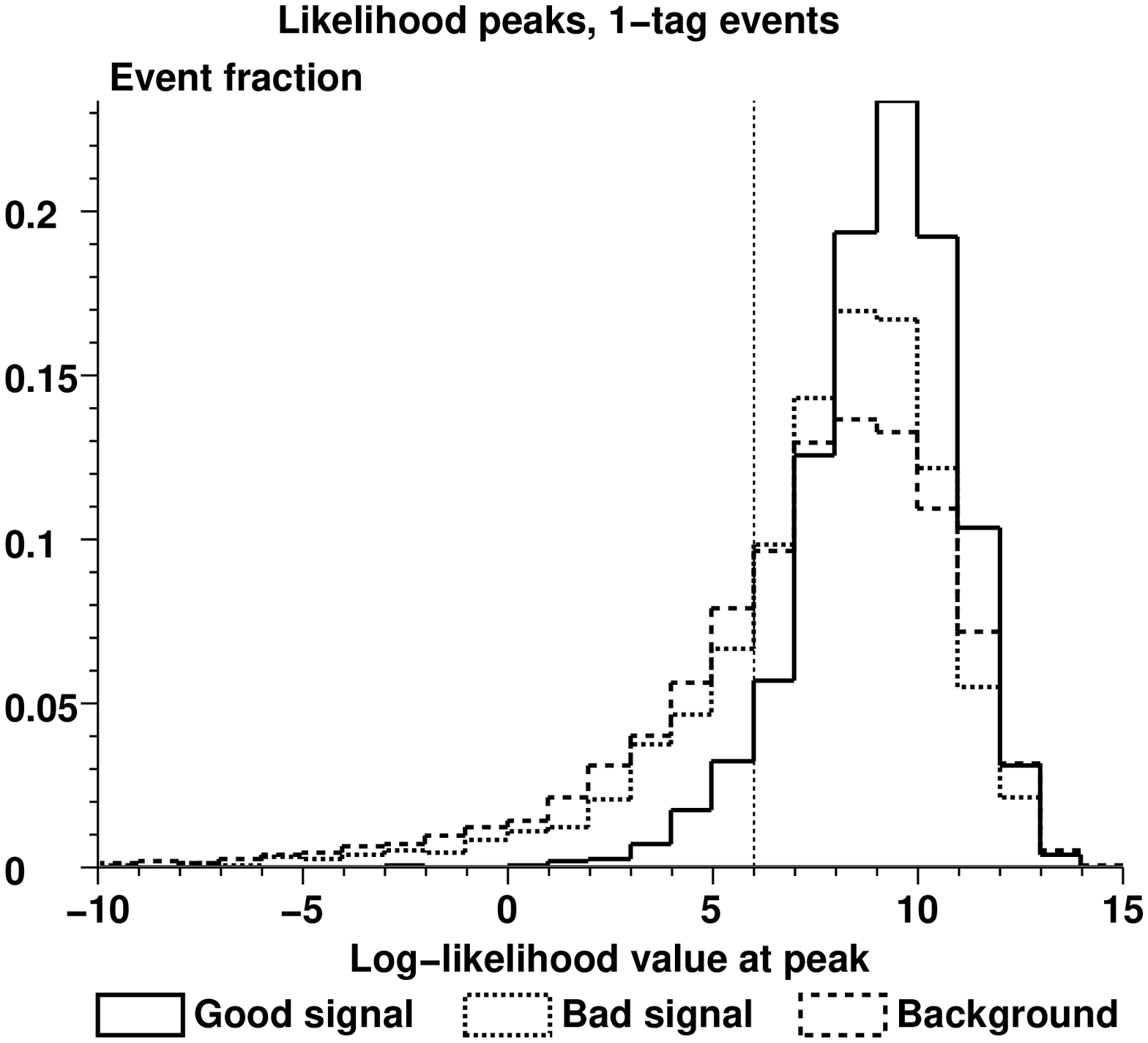}
    \includegraphics*[width=.45\textwidth]{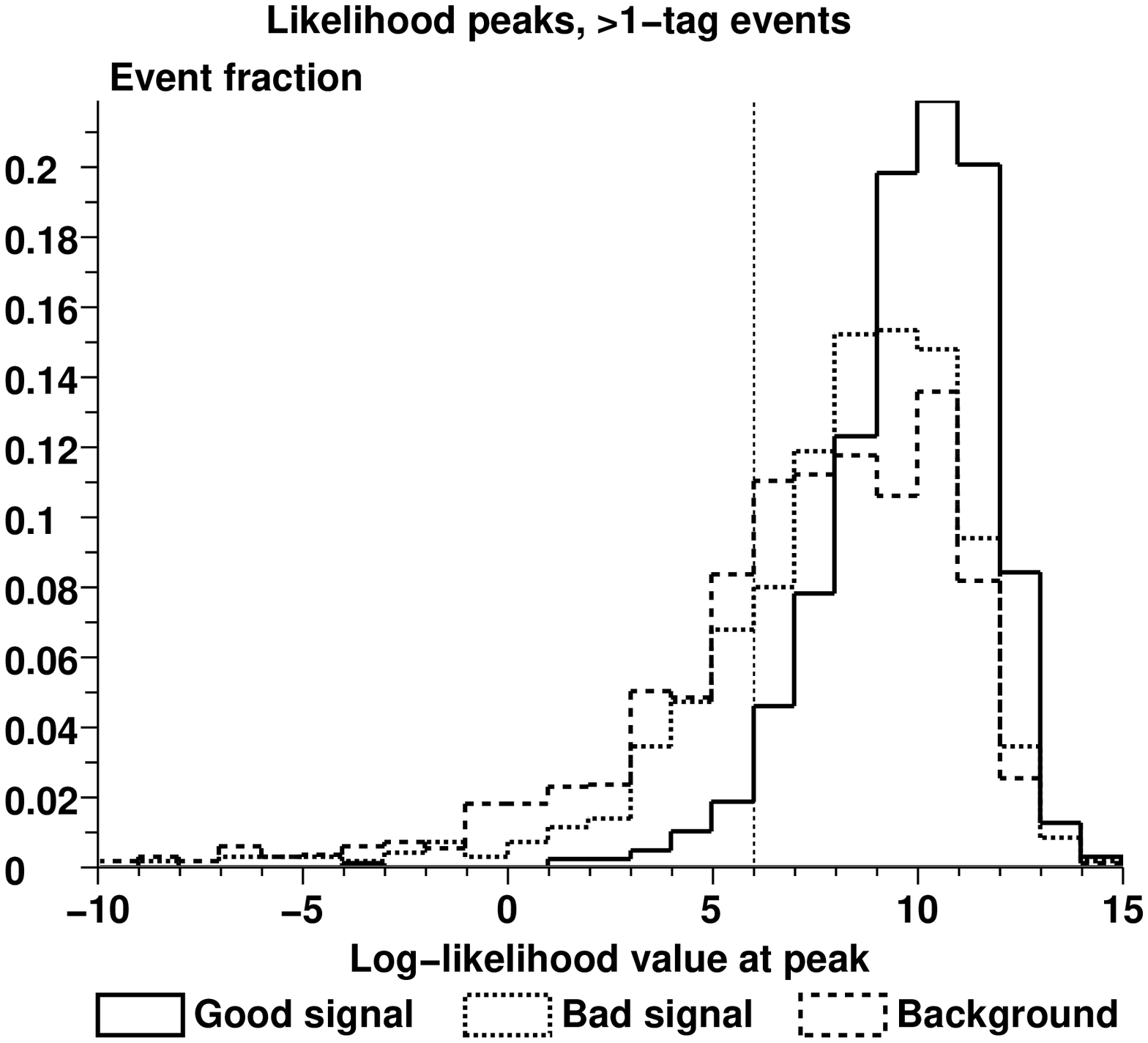}
  \end{center}
\caption{Distributions of the peak value of the log-likelihood for MC
    events, divided into ``good signal'', ``bad signal'', and
  background events.  Left: 1-tag events, right: multiple-tag events. The 
  vertical lines indicate where the likelihood cut is applied.}
\label{fig:peak-likeli}
\end{figure}

We observe that the peaks of the likelihood curves for these ``bad
signal'' events tend to be generally lower than the peaks for
well-behaved $\ttbar$ events. Figure~\ref{fig:peak-likeli} shows the
distribution of the peak value of the likelihood curves for ``good
signal'', ``bad signal'', and background events. We adopt a cut on the
peak value of the likelihood of 6, which retains only the bins
dominated by ``good signal''. Table~\ref{table:likelihood_cut_eff}
shows the efficiency of this cut for ``good signal'' events, ``bad
signal'' events, and background events for $m_t$ = 172 GeV/$c^2$. With
this cut we remove $\sim$22\% of the ``bad signal'' events and
$\sim$29\% of the background events while retaining $\sim$95\% of
``good signal'' events. While this cut reduces the size of our sample,
the overall resolution is significantly improved due to the improved
validity of our assumptions about the sample.

\begin{table}[h]
\caption{Efficiency of the likelihood cut at a value of 6 for $m_t$ =
  172 GeV/$c^2$. The uncertainties shown are the binomial uncertainty
  on the cut, summed appropriately across the different background
  types for the background.}
\label{table:likelihood_cut_eff}
\begin{center}
\begin{tabular}{l@{\hspace{1.0cm}}c@{\hspace{1.0cm}}c}
\hline \hline
Type of event & 1-tag & $>$1-tag \\
\hline
Good signal & 93.6\% $\pm$ 0.4\% & 96.9\% $\pm$ 0.4\% \\
Bad signal  & 76.8\% $\pm$ 0.9\% & 77.5\% $\pm$ 1.5\% \\
Background  & 70.4\% $\pm$ 0.5\% & 68.9\% $\pm$ 1.3\% \\
\hline \hline
\end{tabular}
\end{center}
\end{table}


Our calculation gives us a 2-dimensional joint likelihood 
as a function of $m_t$ and JES. We treat the JES
as a nuisance parameter and eliminate it using the profile likelihood,
i.e., we take the maximum value of the
likelihood along the JES axis for each $m_t$ value. That is:

\begin{eqnarray}
L_{\textrm{prof}}(m_t) = \max_{j \in \JES} L(m_t, j)
\label{eq:mt_profile}
\end{eqnarray}

This gives us a 1-D likelihood curve in $m_t$ only. We then follow the
normal procedure of taking the position of the maximum likelihood as
our reconstructed mass and descending 1/2 unit of log-likelihood from
the peak to determine the estimated uncertainty. Because of
imperfections in our model, these quantities need to be calibrated in
order to obtain a final measured mass and uncertainty.

\section{Test of the Method and Calibration}
\label{sec:calib}

\begin{figure}[htbp]
\begin{center}
\centerline{
\epsfig{file=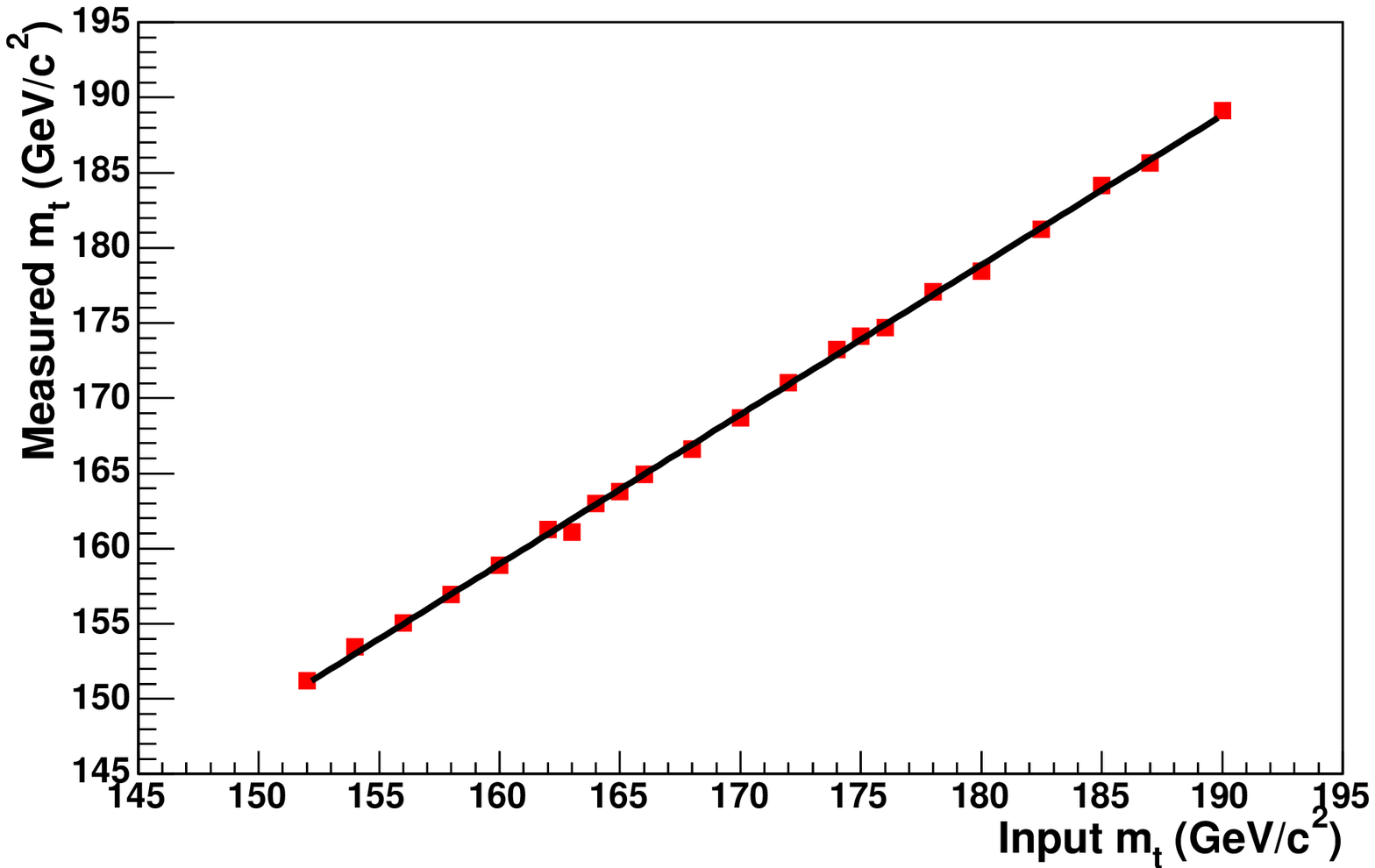, width=.45\textwidth}
\epsfig{file=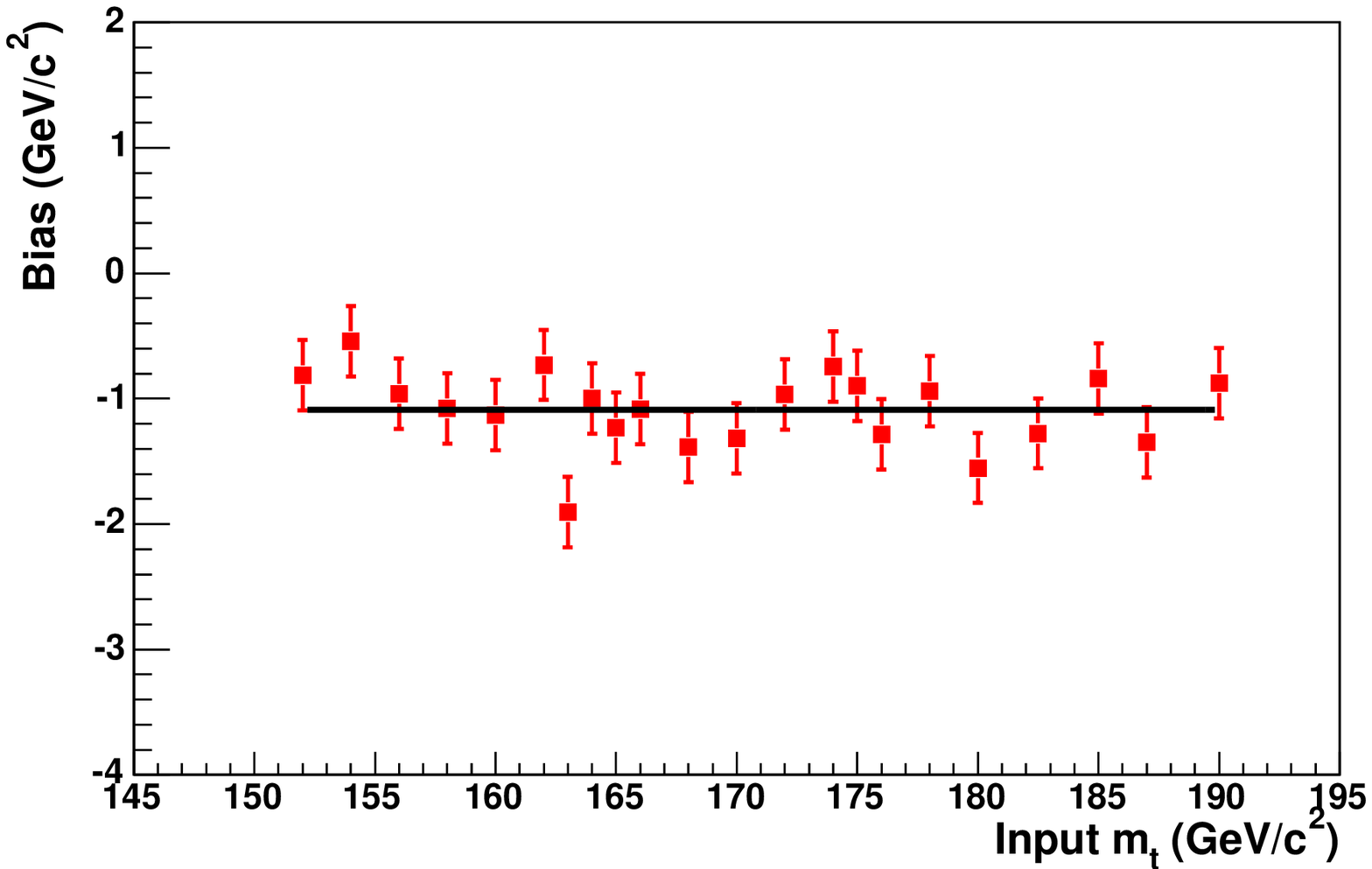, width=.45\textwidth}
}
\centerline{
\epsfig{file=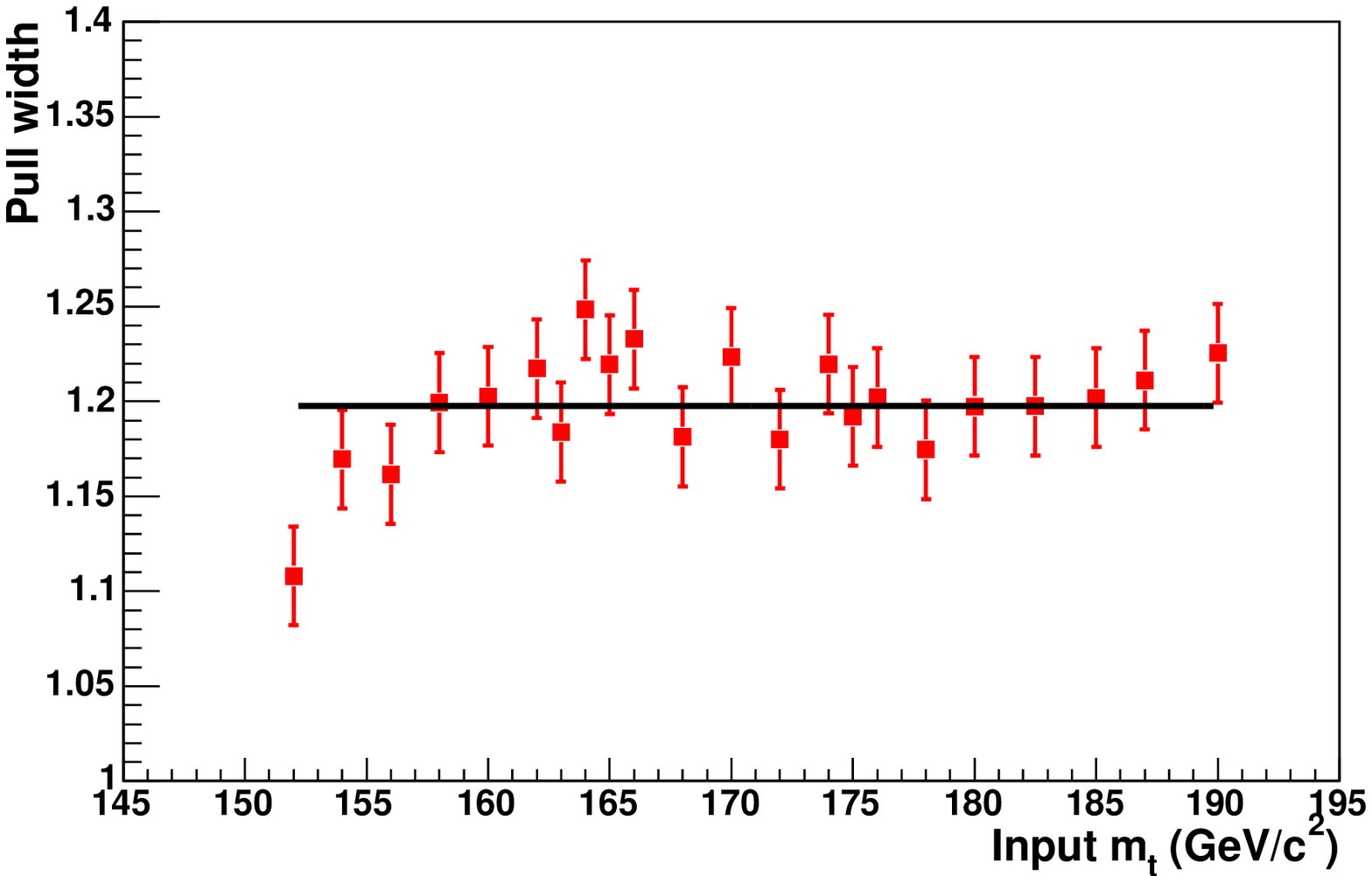, width=.45\textwidth}
}
\caption{Pseudoexperiment results using fully simulated signal and
  background events after applying a likelihood cut, with a mean of
  303 events for each PE.  For these samples, JES is fixed at its
  nominal value of 1. Top left: reconstructed vs. input top quark
  mass; top right: bias vs. input top quark mass; bottom: pull width
  vs. input top quark mass.}
\label{fig:pe_results_cut}
\end{center}
\end{figure}

We test our method using MC samples of fully simulated and
reconstructed $t\bar{t}$ events and the background samples described
in Section~\ref{sec:sample}. We construct pseudoexperiments (PEs) from
the MC samples with an average total number of events equal to the
number observed in the data. As shown in Table~\ref{ev_back} we
observe 371 events, of which 70.3 $\pm$ 16.5 are expected to be
background. After applying the likelihood cut efficiencies for signal
and background simulated events, we expect a total of 303 events,
which is the number of events that we use in our PEs. The number of
each type of event (signal and each background type) is
Poisson-fluctuated about its expected contribution to the total.  We
perform 2000 PEs for each signal top quark mass value and compute the
resulting average reconstructed mass, bias, expected statistical
uncertainty, and pull width. Figure~\ref{fig:pe_results_cut} shows the
reconstructed mass, bias, and pull width versus the input top quark
mass, where the bias is defined as the difference between the true
mass $m_{\textrm{true}}$ and the reconstructed mass $m_{\textrm{rec}}$
and the pull width is the width of the distribution of
$(m_{\textrm{rec}} - m_{\textrm{true}})/(\sigma_m)_{\textrm{rec}}$ in
individual PEs, where $(\sigma_m)_{\textrm{rec}}$ is the estimated
uncertainty.  The output mass is a linear function of the input mass
with a slope very close to 1; the mass bias and the pull width are
independent of the input top quark mass. The non-zero bias and
non-unit pull width are due to the presence of events not modeled in
our effective propagator model (``bad signal'' events and background)
in our analysis; if we run PEs on ``good signal'' events only, we
obtain a bias and average pull width consistent with 0 and 1,
respectively.

We use the observed slope, bias, and pull width to calibrate our
method. Specifically, the reconstructed mass is calibrated by
correcting for the measured bias and slope, and the estimated
uncertainty is calibrated by correcting for the measured pull width
and slope. We perform our calibrations in terms of $\Delta m_t \equiv
m_t - 172$ GeV/$c^2$; our overall calibration formulae to obtain a
calibrated mass $\Delta m_{\textrm{cal}}$ and mass uncertainty
$(\sigma_m)_{\textrm{cal}}$ in terms of the observed mass $\Delta
m_{\textrm{obs}}$ and uncertainty $(\sigma_m)_{\textrm{obs}}$ are:

\begin{eqnarray}
\Delta m_{\textrm{cal}} & = & (\Delta m_{\textrm{obs}} - c_2) / c_1, \\
\label{eq:dm_calib}
(\sigma_m)_{\textrm{cal}} & = & (\sigma_m)_{\textrm{obs}} \times c_3 /
c_1,
\label{eq:dm_sigma_calib}
\end{eqnarray}

where $c_1$, $c_2$, and $c_3$ are the slope in the upper-left plot,
the constant in the upper-right plot, and the constant in the bottom
plot of Fig.~\ref{fig:pe_results_cut}, respectively. Using the fits
shown in these plots, we obtain $c_1 = 0.995 \pm 0.006$, $c_2 = -1.09
\pm 0.06$ GeV/$c^2$, and $c_3 = 1.20 \pm 0.01$.

\begin{figure}[htbp]
\begin{center}
\centerline{
\epsfig{file=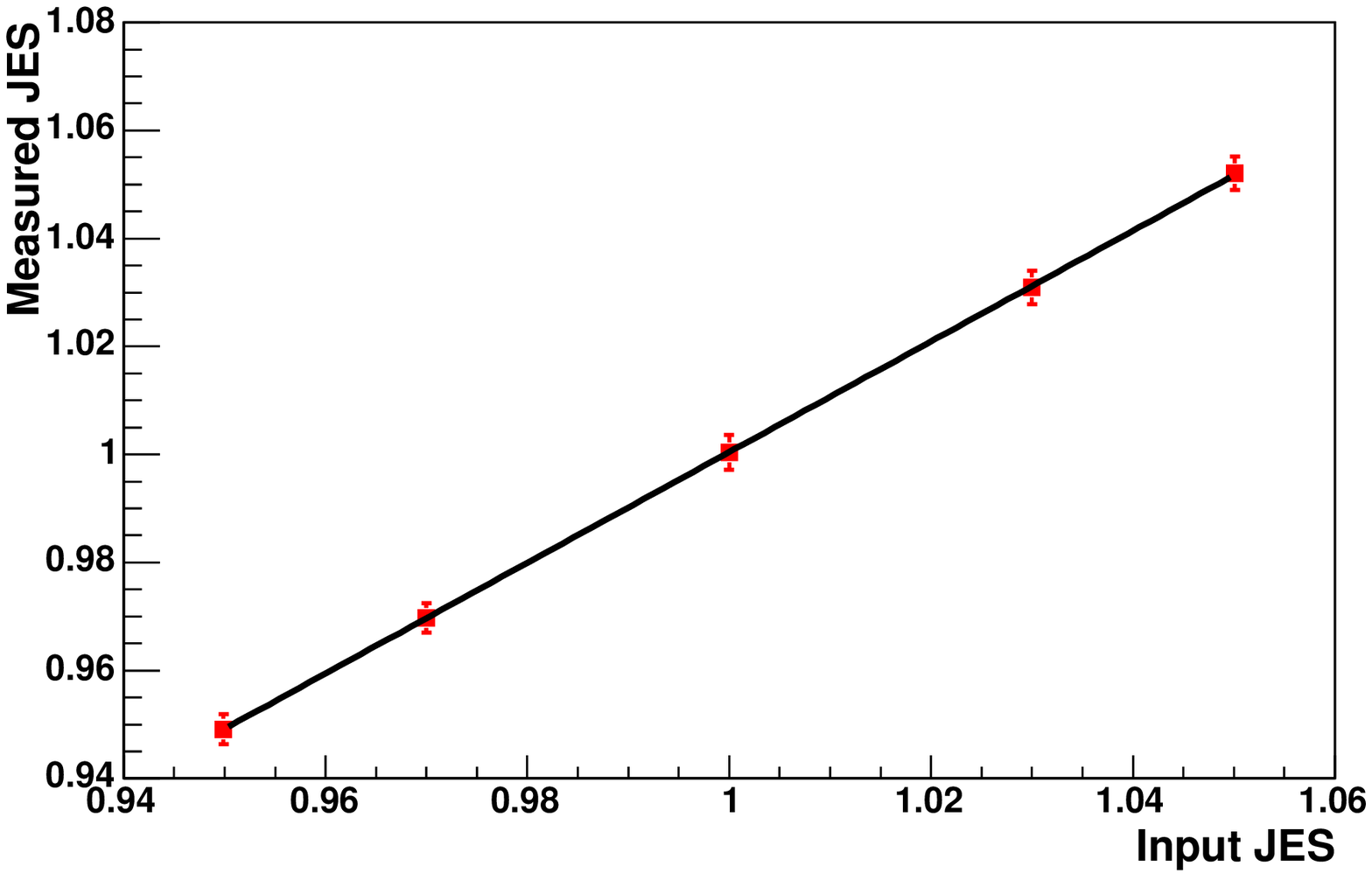, width=.45\textwidth}
\epsfig{file=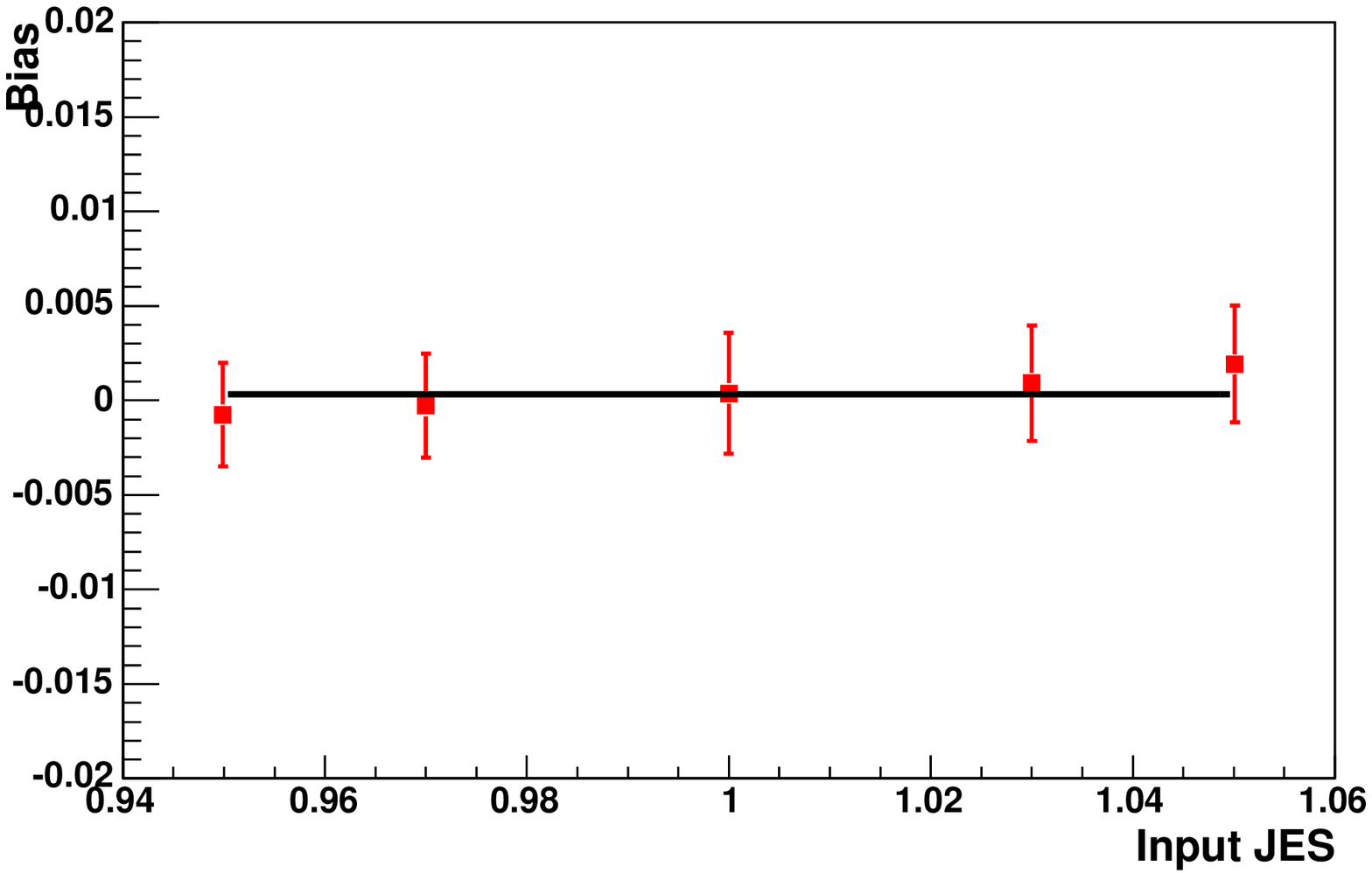, width=.45\textwidth}
}
\centerline{
\epsfig{file=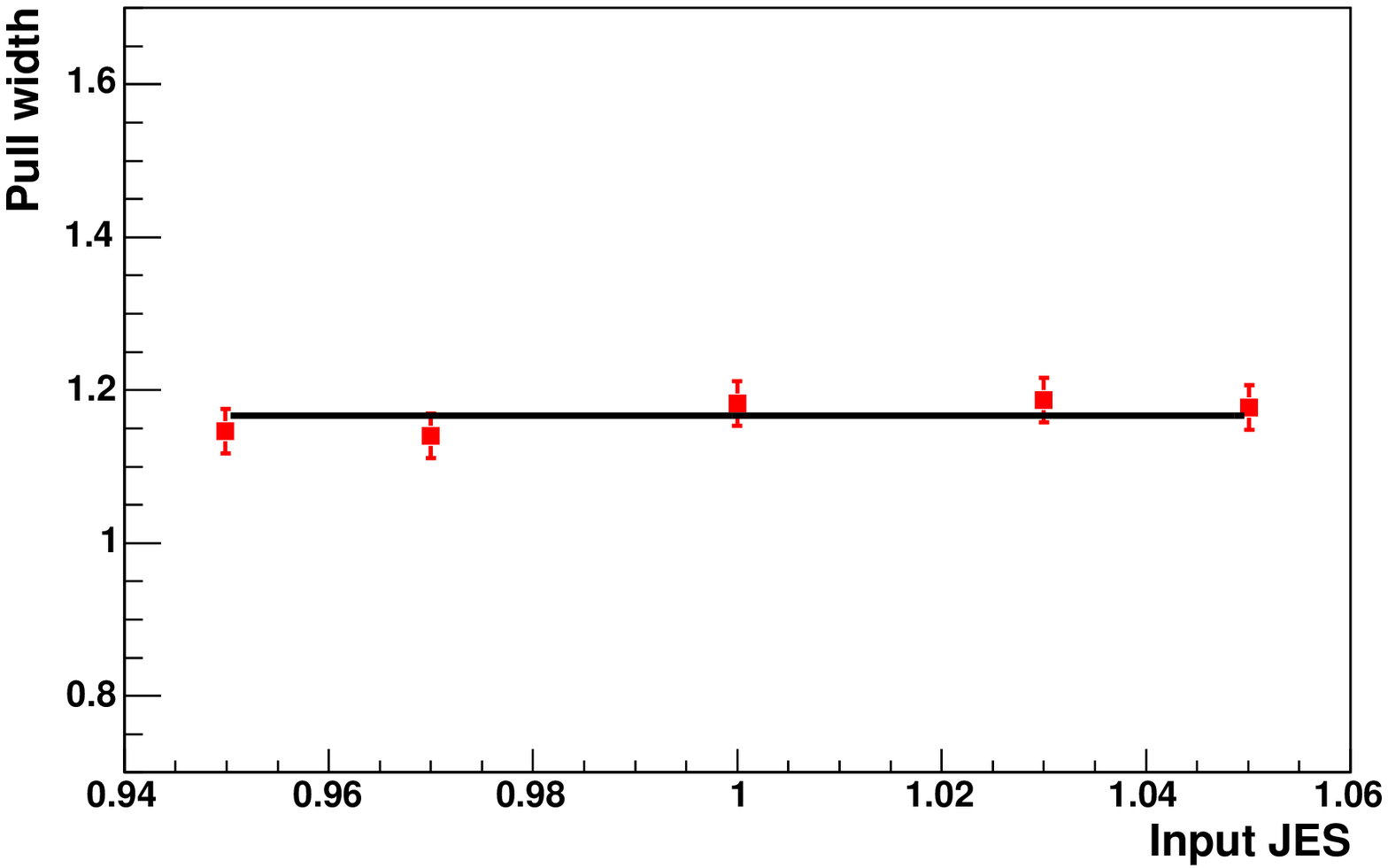, width=.45\textwidth}
}
\caption{JES pseudoexperiment results using fully simulated signal and
  background events after applying a likelihood cut, with a mean of
  303 events for each PE. The top quark mass here is fixed at $m_t =
  170$ GeV/$c^2$. Top left: reconstructed vs. input JES; top right:
  JES bias vs. input JES; bottom: JES pull width vs. input JES.}
\label{fig:jes_results_cut}
\end{center}
\end{figure}

The 2-D likelihood method measures JES {\it in situ} in the $t\bar t$
sample. To ensure that this method correctly handles events where the
JES is not necessarily equal to its nominal value, we also check
simulated samples where the JES has been shifted from its nominal
value of unity. Specifically, we use four different JES shifts: JES =
0.95, 0.97, 1.03, and 1.05, to obtain a calibration for our JES
measurement in the same way that we calibrate our $m_t$ measurement
above. Figure~\ref{fig:jes_results_cut} shows the reconstructed JES,
JES bias, and JES pull width versus the input JES for $m_t = 170$
GeV/$c^2$.  The non-unity pull width for JES is due to the same
origin as the non-unity pull width for the top mass. We use these
results to obtain our calibration for the reconstructed JES. We
perform our calibrations in terms of $\Delta(\JES)
\equiv \JES - 1$, yielding the final formulae for our calibrated $\Delta(\JES)$,
$\Delta(\JES)_{\textrm{cal}}$, and JES uncertainty
$(\sigma_{\JES})_{\textrm{cal}}$ in terms of the observed value,
$\Delta(\JES)_{\textrm{obs}}$, and uncertainty
$(\sigma_{\JES})_{\textrm{obs}}$:

\begin{eqnarray}
\Delta(\JES)_{\textrm{cal}} & = & (\Delta(\JES)_{\textrm{obs}} - c_5) / c_4 \\
(\sigma_{\JES})_{\textrm{cal}} & = & (\sigma_{\JES})_{\textrm{obs}}
\times c_6 / c_4,
\label{djes_calib}
\end{eqnarray}

where $c_4$, $c_5$, and $c_6$ are the slope in the upper-left plot,
the constant in the upper-right plot, and the constant in the bottom
plot of Fig.~\ref{fig:jes_results_cut}, respectively. Using these results,
we obtain $c_4 = 1.03 \pm 0.04$, $c_5 = 0.0003 \pm 0.0013$, and $c_6 =
1.17 \pm 0.01$.

\begin{figure}[htbp]
\centerline{
\epsfig{file=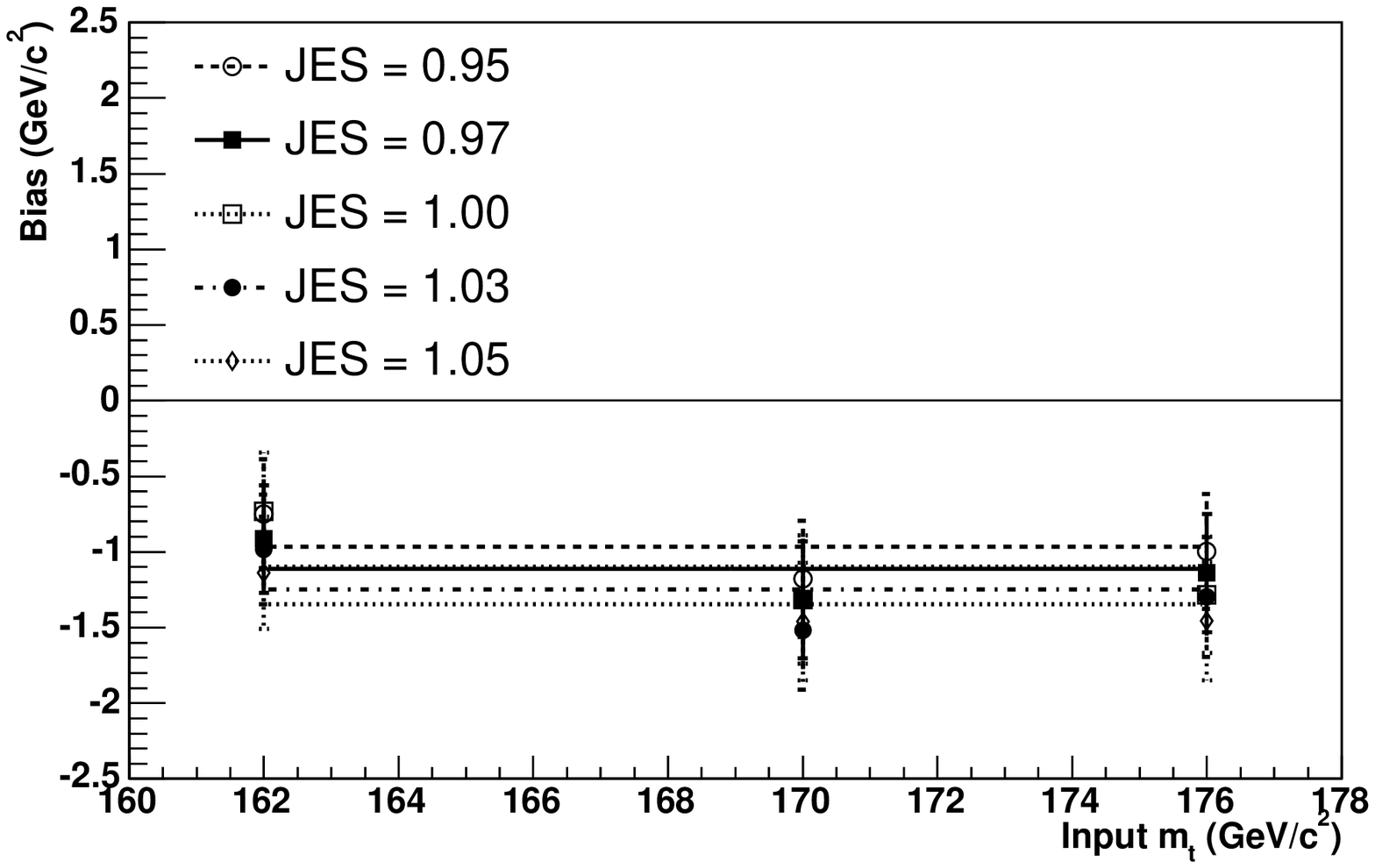, width=.45\textwidth}
\epsfig{file=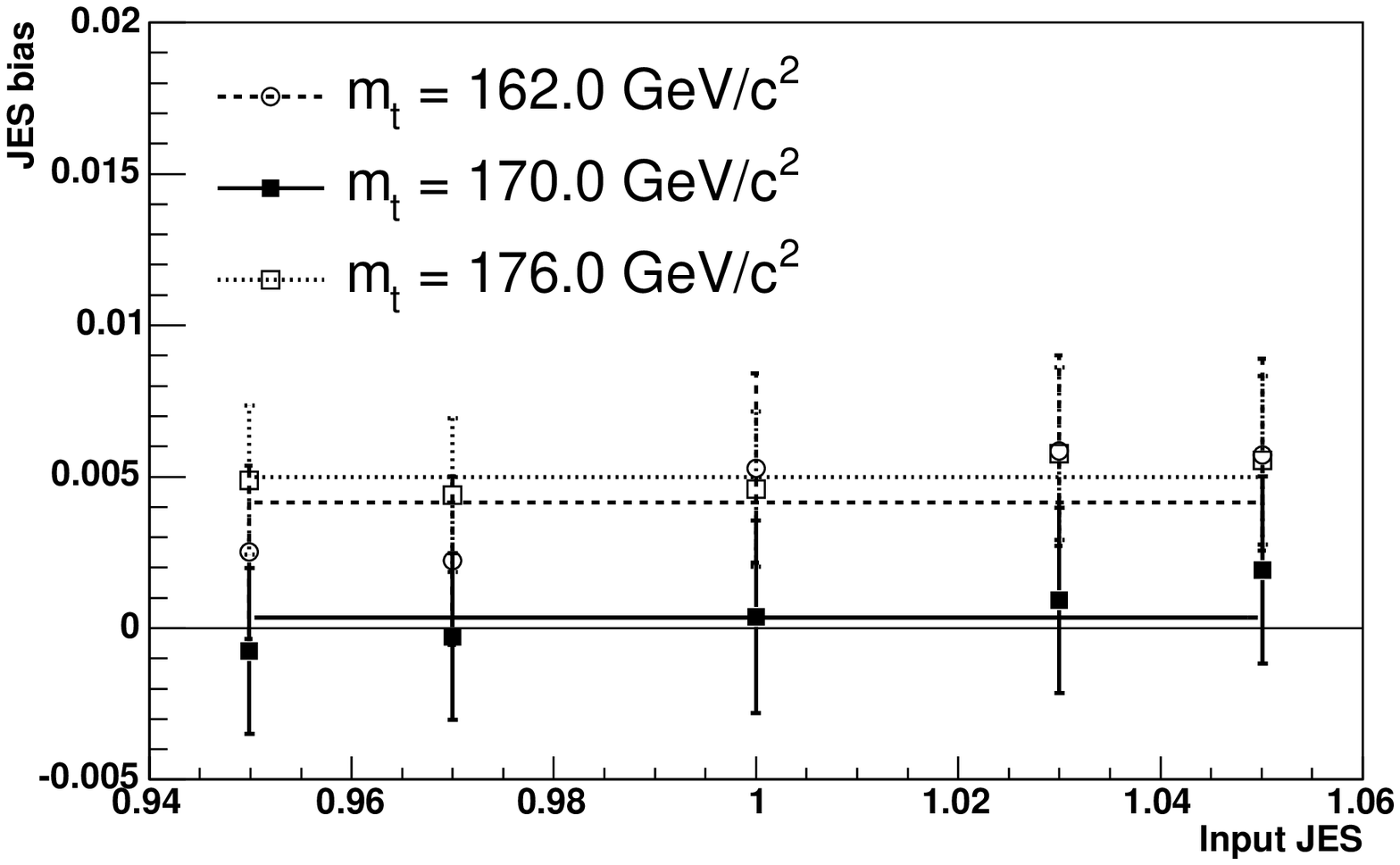, width=.45\textwidth}
}
\centerline{
\epsfig{file=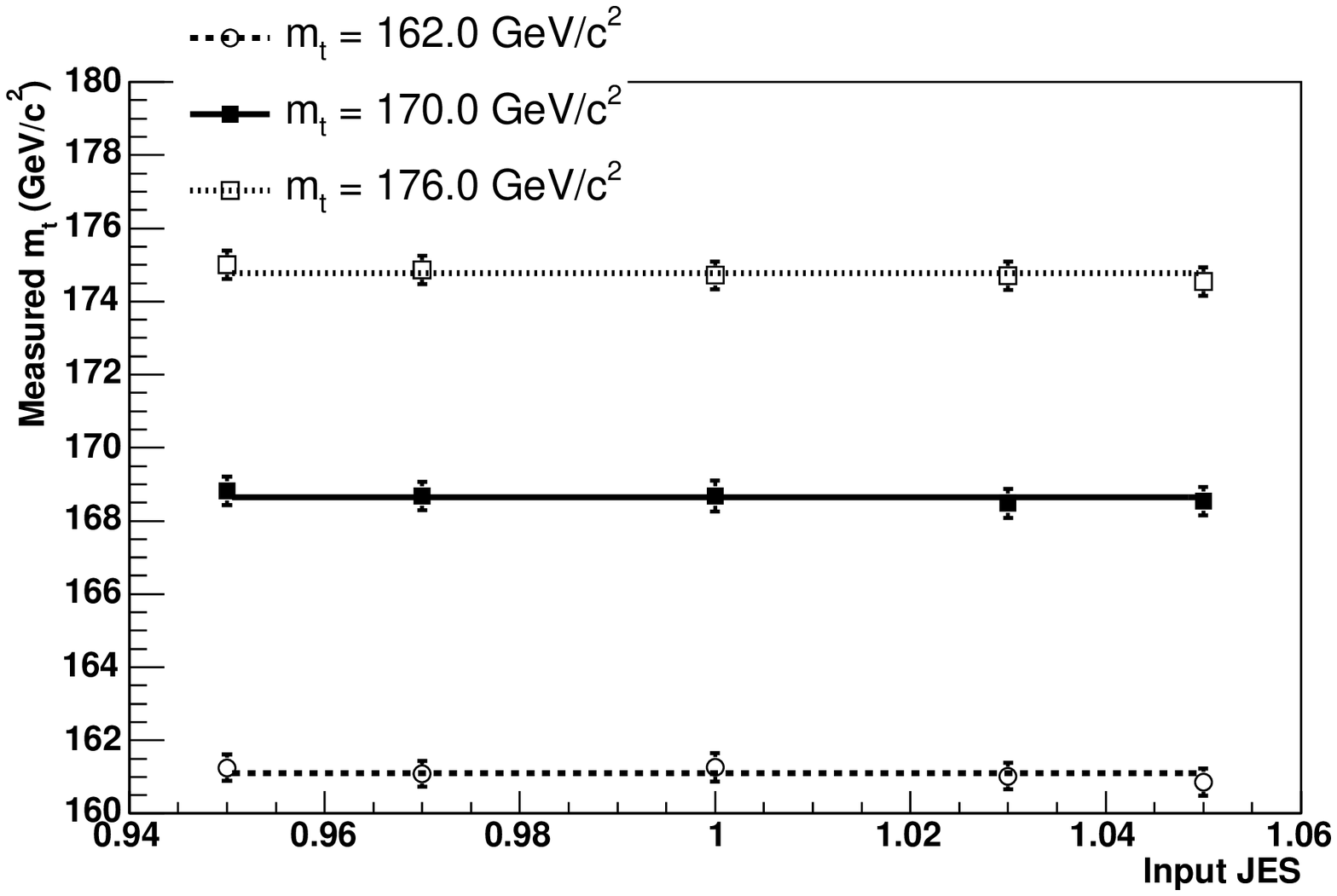, width=.45\textwidth}
}
\caption{Results of studying samples with shifted JES values. Top
  left: Mass bias vs. input top quark mass at the five different JES
  points; top right: JES bias vs. input JES at the three
  different mass points; bottom: reconstructed top quark mass
  vs. input JES for the three different top quark masses.}
\label{figure:jes-lin}
\end{figure}

Since the $m_t$ and JES calibration parameters are derived for JES
fixed at 1 and $m_t$ fixed at 170 GeV/$c^2$, respectively, we also need to
ensure that they do not vary for different values of $m_t$ and
JES. Figure~\ref{figure:jes-lin} shows the results of these
studies. We note that the JES and $m_t$ slope and bias do not
noticeably change for different $m_t$ and JES inputs. The plot on the
bottom also shows that the reconstructed top quark mass is very stable with
respect to the input JES, showing that our procedure of independent
calibration of the two variables is valid.

\section{Systematic Uncertainties}
\label{sec:systs}

We discuss the systematic uncertainties on our measurement in this
section. Most systematics are evaluated using a general procedure
where we shift a given quantity by $\pm 1 \sigma$ in the signal MC
sample and/or in the background samples, perform PEs to measure a top
quark mass on the shifted sample, and use the resulting shift in the
measured top quark mass as our systematic uncertainty. In this paper,
we have rounded all of the final systematic uncertainties to a
precision of 0.1 GeV/$c^2$; although some systematics are known to a higher
precision than this, some are not.

\begin{itemize}
\item \textbf{Calibration:} In calibrating our final result, we
use the bias and slope constants $c_1$, $c_2$, and $c_3$  as described in
Section~\ref{sec:calib}. The uncertainty on these constants is a
source of systematic uncertainty in our measurement. The uncertainty
in $c_1$ is the major source of our final quoted uncertainty of 0.1
GeV/$c^2$.

\item \textbf{Residual JES:} Although the 2-D measurement is designed
to capture any changes in the JES, we assume a constant factor for the
jet energy scale, whereas the jet energy systematic uncertainties
depend on the jet $p_T$ and $\eta$. Furthermore, the JES uncertainties
are composed of the sum of several different potential sources, each
of which may vary differently. To evaluate potential systematic
uncertainties due to this assumption, we shift the jet energies by one
standard deviation for each source of systematic uncertainty
(corrections for relative response of different sections of the
calorimeter, the absolute corrections in cone with $\Delta R = 0.4$,
out of cone correction details, and other minor sources as described
in Ref.~\cite{TMT}). We perform these tests shifting only the
signal, and also shifting the signal together with the $W + b\bar{b}$
background, and take the greater of the resulting differences. The
resulting shifts are added in quadrature to obtain our residual JES
systematic uncertainty of 0.5 GeV/$c^2$.

\item \textbf{$b$-JES:} We have assumed that the JES is the
same for all jets. However, there is an additional uncertainty arising
from relative differences between $b$- and light-quark jets. (Note
that the jet systematic uncertainties are predominantly determined
using light jets.) We identify three sources of uncertainty: one due
to the uncertainty in the semileptonic decay ratio, which we estimate
by varying this ratio by $\pm 1 \sigma$; one due to the uncertainty in
the $b$-fragmentation modeling, which we evaluate by varying the
parameters used in the Bowler fragmentation model~\cite{bfrag} in the
{\sc pythia} Monte Carlo generator, using two different sets of
parameters derived from SLD and LEP results~\cite{bfrag_param}; and
one due to uncertainty in the calorimeter response for $b$-jets
compared to light-quark jets. Because the calorimeter response is determined in light-jet samples, the different charged particle fraction and momentum spectrum in $b$-jets could result in a different response.  We evaluate this uncertainty by checking the effect of the calorimeter corrections in simulated light jets and $b$-jets separately.  Then we propagate this difference in the response by
shifting the $E_T$ of jets identified as $b$-jets in the $t \bar t$ Monte Carlo
sample. The three corrections to the $b$-jet energy scale yield uncertainties of 0.1
GeV/$c^2$, 0.3 GeV/$c^2$, and 0.1 GeV/$c^2$, respectively, for an
overall uncertainty of 0.3 GeV/$c^2$.

\item \textbf{Generator:} We evaluate a systematic due to the
MC generator used by comparing the results from {\sc herwig} and {\sc
pythia} $\ttbar$ samples. We take the resulting difference of 0.6
GeV/$c^2$ as our systematic uncertainty. There is also a potential
systematic uncertainty for color reconnection effects not included
here~\cite{cr}; current studies suggest that these may not
significantly increase our total systematics.

\item \textbf{ISR and FSR:} Systematic errors due to initial-state
radiation (ISR) and final-state radiation (FSR), where additional
gluons are radiated, are evaluated using MC samples.  A control sample
of Drell-Yan events allows us to compare the dilepton invariant mass
spectrum in data and MC and estimate the uncertainty on the parameters
controlling the radiation~\cite{TMT}.  Those parameters have been
varied by the estimated uncertainty to study the effect on the top
mass measurement.  In this case the uncertainty on the measured top
mass shift is larger than the shift itself, so we use the uncertainty
of 0.3 GeV/$c^2$ as our quoted uncertainty.

\item \textbf{PDFs:} We evaluate the systematic uncertainty due to
the parton distribution functions (PDFs) used in the matrix element
integration by comparing different PDF sets (CTEQ5L~\cite{cteq} and
MRST72~\cite{mrst}), varying $\alpha_s$, and varying the eigenvectors
of the CTEQ6M PDFs. The final uncertainty is defined by the sum of the
eigenvector uncertainty in quadrature and the $\alpha_s$
difference. The CTEQ-MRST difference in our case is negligible. The
result is an overall uncertainty of 0.4 GeV/$c^2$.

\item \textbf{Background:} There are several uncertainties
associated with our background subtraction. First is the uncertainty
due to uncertainty on the overall background fraction. The second
source is the uncertainty in the background composition, which we
assess by setting the background to, in turn, 100\% $W + b\bar{b}$, $W
+ c\bar{c}/c$, $W$ + light, or QCD background and taking the largest
resulting shift as our uncertainty. Third, the uncertainty associated
with our average background likelihood
$\overline{L_{\textrm{bg}}}(m_t, \JES)$ as described in
Section~\ref{sec:tot-likeli}; to evaluate this uncertainty, we divide
the sample into two disjoint subsamples (one with only electrons, and
one with only muons), build the average background likelihood curve
from one subsample, and measure the top quark mass using the other
subsample. Finally, we account for uncertainties due to the $Q^2$
scale used by the background MC generator. The resulting systematic
uncertainties are 0.3 GeV/$c^2$, 0.4 GeV/$c^2$, 0.3 GeV/$c^2$, and 0.2
GeV/$c^2$, respectively.

\item \textbf{Lepton $p_T$:} To account for the 1\% uncertainty on
the measured lepton $p_T$, we apply our method to samples where 
the lepton $p_T$
has been shifted by this amount, resulting in an uncertainty on the
top quark mass of 0.1 GeV/$c^2$.

\item \textbf{Permutation weighting:} We account for a potential
systematic for the tagging probabilities used to weight our
permutations (the $w_i$ factors in Eq.~\ref{eq:prob_integral}), 
since these are derived from fits to the tagging
probabilities measured in data. We estimate that the predominant
source of uncertainty in this estimate is the ratio of charm tags to
$b$-tags, which is nominally 22\%. We vary this by its relative
uncertainty of 15\% and measure the resulting difference, which is
negligible.

\item \textbf{Pileup:} We consider two sources of uncertainty due to
multiple $\ppbar$ interactions. First, we consider the fact that the
number of interactions in our Monte Carlo samples is not equal to the
number observed in the data. To estimate this effect, we divide our
Monte Carlo samples into subsamples with different number of
interactions in the event, examine the slope of the resulting measured
top quark mass as a function of the number of interactions, and
multiply this by the difference in the number of interactions between
Monte Carlo events and data events. Second, we consider the modeling
of the additional interactions in an event. Our current model is
derived from minimum bias events, so we consider the possibility that
it does not correctly model $\ttbar$ events. For this purpose, we
compare the observed jet response as a function of the number of
vertices in $\ttbar$ Monte Carlo events and minimum bias data and use
the resulting difference to obtain a systematic uncertainty. We take
the larger of these two sources, 0.2 GeV/$c^2$, as our systematic
uncertainty.

\item \textbf{Gluon fraction:} {\sc herwig} and {\sc pythia} are
both leading-order MC generators, so $\ttbar$ events in these samples
are approximately 95\% produced from $q\bar{q}$ annihilation and 5\%
produced from $gg$ fusion. However, NLO expectations are closer to ($15
\pm 5)\% ~gg$ production. To check for a potential systematic due to this
effect, we run PEs where $q\bar{q}$ and $gg$ events are reweighted so
that the $q\bar{q}$ weights sum to 0.80 and the $gg$ events to 0.20
(using the maximal $gg$ percentage to be conservative) and use the
resulting shift of 0.3 GeV/$c^2$ as our uncertainty.

\end{itemize}

Table~\ref{table:syst} summarizes our final list of systematic
uncertainties.

\begin{table}[h]
\caption{Total list of systematic uncertainties.}
\label{table:syst}
\begin{center}
\begin{tabular}{r@{\hspace{1.0cm}}c}
\hline \hline
Systematic source & Systematic uncertainty (GeV/$c^2$) \\
\hline
Calibration                & 0.1 \\
Residual JES               & 0.5 \\
$b$-JES                    & 0.3 \\
MC generator               & 0.6 \\
ISR and FSR                & 0.3 \\
PDFs                       & 0.4 \\
Background: fraction       & 0.3 \\
Background: composition    & 0.4 \\
Background: average shape  & 0.3 \\
Background: $Q^2$          & 0.2 \\
Lepton $p_T$               & 0.1 \\
Pileup                     & 0.2 \\
Gluon fraction             & 0.3 \\
\hline
Total                      & 1.2 \\
\hline \hline
\end{tabular}
\end{center}
\end{table}

\section{Results}
\label{sec:result}

\begin{figure}[htbp]
\centerline{
\epsfig{file=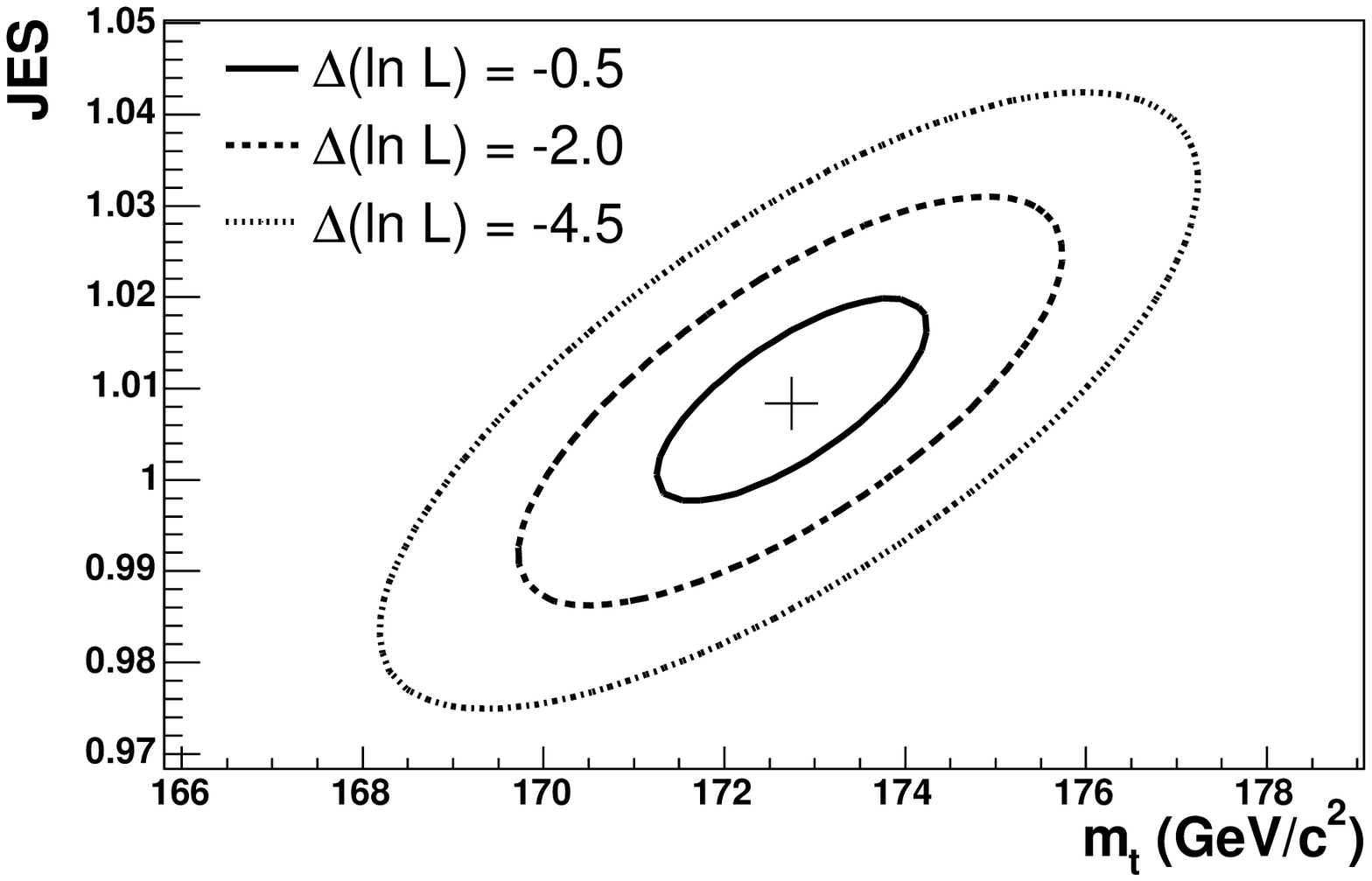,width=.45\textwidth}
\epsfig{file=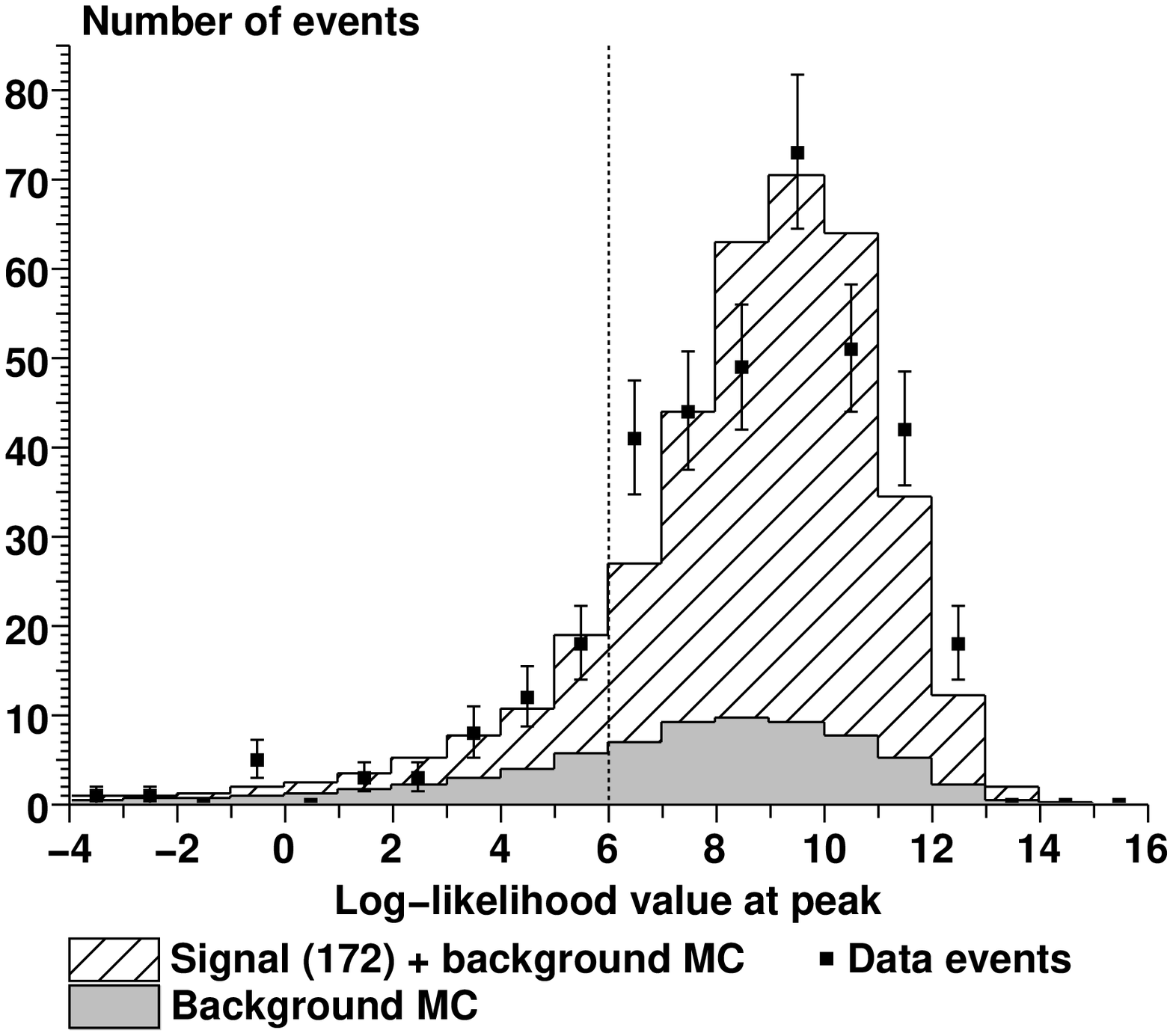,width=.45 \textwidth}
}
\caption{Left: Measured 2-D likelihood on the data events.  The
  plot shows the contours corresponding to a 1-$\sigma$,
  2-$\sigma$, and 3-$\sigma$ uncertainty (assuming Gaussian behavior) 
  in our measurement.  The
  calibration derived from MC events has been applied to both axes.
  The marker shows the point of maximum
  likelihood. Right: Likelihood peak position of the individual likelihood
  curves for data and MC events. The dashed line indicates the
  likelihood cut of 6 employed.}
\label{fig:data_results}
\end{figure}

In the data we find a total of 318 events which pass all of our
selection requirements (including the likelihood peak value cut), of which
237 have exactly 1 tag and 81 have more than 1 tag.
We combine the likelihoods for the 1-tag and $>$1-tag subsamples using
the formula described in Eq.~\ref{eq:final_bg_subtraction} and
then combine the two subsamples. After obtaining this total
likelihood, we use the profile likelihood method introduced in
Section~\ref{sec:tot-likeli} to extract a top quark mass value.  That
mass value is then corrected using the calibration procedure to obtain
a mass value of $172.7 \pm 1.8$~GeV/$c^2$.  The left plot in
Fig.~\ref{fig:data_results} shows the resulting 2-D likelihood
contours after calibration for 1-$\sigma$, 2-$\sigma$, and 3-$\sigma$
uncertainties, assuming that they have Gaussian distributions.

To validate the likelihood cut used in our
procedure, we compare distributions of the event peak likelihoods in
data and MC events. The K-S confidence level for these two
distributions is 93.8\%, indicating a very good agreement between the
data and MC simulations.

\begin{figure}[htbp]
\centerline{
\epsfig{file=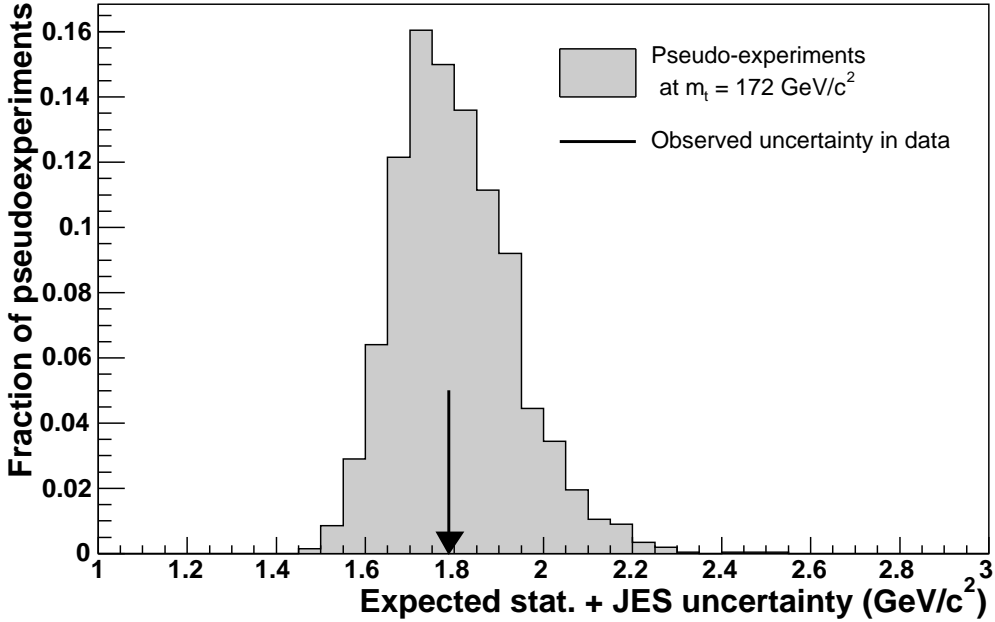,width=.9\textwidth}
}
\caption {Expected statistical uncertainty (including uncertainty due
to JES) on the top mass from the 2-D profile
likelihood method, derived from MC events with $m_t = 172$
GeV/$c^2$. The black arrow indicates the uncertainty in the data
measurement. All uncertainties have been scaled by the average pull
width of 1.198.  They are also corrected by 1/0.995 to account for a measured response slope slightly different from 1. }
\label{fig:error}
\end{figure}

This result combines the statistical uncertainty and the uncertainty
due to JES. To separate these two contributions, we fix the JES value
to be 1.0 and evaluate the uncertainty in the resulting 1-D likelihood
as a function of $m_t$. This yields an uncertainty of $1.2$
GeV/$c^2$. We conclude that the remaining uncertainty of $1.3$
GeV/$c^2$ is due to the JES.
The expected statistical + JES
uncertainty from MC events at a top quark mass of 172 GeV/$c^2$ peaks
at 1.8 GeV/$c^2$, in good agreement with the measured uncertainty in the
observed event sample of 1.8 GeV/$c^2$; 50\% of pseudoexperiments
show a smaller uncertainty than that measured in the data. The
distribution of the expected uncertainties is shown
in Fig.~\ref{fig:error}. The additional systematic uncertainty on
the measured top mass discussed in Section~\ref{sec:systs} is 1.2
GeV/$c^2$, yielding a final result of:

\begin{center}
$m_t = 172.7 \pm 1.2~\textrm{(stat.)} \pm 1.3~\textrm{(JES)} 
\pm 1.2~\textrm{(syst.)} ~\textrm{GeV}/c^2$
\end{center}

From the 2-D likelihood, we can also obtain a JES measurement by using
the profile likelihood technique to eliminate the $m_t$ axis and
applying the calibration described previously. This yields a
measurement of $\JES = 1.008 \pm 0.013$, indicating that the JES is
well within its expected value. Note that the profile likelihood
measurement technique does not impose any prior expectations on our
JES value.

\section{Conclusion}
\label{sec:summary}

We have used 1.9 fb$^{-1}$ of data at the Tevatron to measure the mass
of the top quark using the lepton + jets topology. Our analysis uses a
modified matrix element integration technique.  To date, analyses
using this technique have made simplifying kinematic assumptions for
the purposes of computational tractability. The method described in
this paper includes the first attempt to compensate directly for these
assumptions. Our measured top quark mass with 318 events passing all
our selection criteria is:
 
$$m_t = 172.7 \pm 1.8~\textrm{(stat. + JES)} \pm 1.2~
\textrm{(syst.)}~\textrm{GeV}/c^2$$
or combining statistical and systematic uncertainties (assuming
Gaussian behavior)
 
$$m_t = 172.7 \pm 2.1~\textrm{(total)}~\textrm{GeV}/c^2. $$ 

Our model at
the moment does not take into account events where a jet from top
quark decay is missing or is replaced by a jet from the parton
shower. Proper treatment of these ``bad signal'' events should
help improve the measurement.

\begin{acknowledgments}
We thank the Fermilab staff and the technical staffs of the
participating institutions for their vital contributions. This work
was supported by the U.S. Department of Energy and National Science
Foundation; the Italian Istituto Nazionale di Fisica Nucleare; the
Ministry of Education, Culture, Sports, Science and Technology of
Japan; the Natural Sciences and Engineering Research Council of
Canada; the National Science Council of the Republic of China; the
Swiss National Science Foundation; the A.P. Sloan Foundation; the
Bundesministerium f\"ur Bildung und Forschung, Germany; the Korean
Science and Engineering Foundation and the Korean Research Foundation;
the Science and Technology Facilities Council and the Royal Society,
UK; the Institut National de Physique Nucleaire et Physique des
Particules/CNRS; the Russian Foundation for Basic Research; the
Ministerio de Ciencia e Innovaci\'{o}n, and Programa
Consolider-Ingenio 2010, Spain; the Slovak R\&D Agency; and the
Academy of Finland.
\end{acknowledgments}


\begin{thebibliography}{999}
\bibitem{TMT} A. Abulencia {\it et al.} (CDF Collaboration),
 Phys. Rev. D \textbf{73}, 032003 (2006).  Note that we use a
  procedure slightly different from the one described in this article;
  we use samples in which the ISR and FSR have been shifted together
  by a $\pm 1 \sigma$ variation. The determination of the $1 \sigma$
  uncertainty is the same as described in this article.
\end{thebibliography}
\end{document}